\documentclass[iop,apj]{emulateapj}
\slugcomment{{\sc Accepted to ApJ:} April 27, 2015}

\usepackage{amsmath}
\usepackage{amssymb}
\usepackage{graphicx,graphics}
\usepackage[english]{babel}
\usepackage{natbib}
\usepackage{color}

\begin{document}

\title{On the Influence of Minor Mergers on the Radial Abundance Gradient in Disks of Milky Way-like Galaxies}

\author{
       Igor A. Zinchenko  \altaffilmark{1,2},
       Peter Berczik  \altaffilmark{3,1,2},
       Eva K. Grebel \altaffilmark{2},
       Leonid S. Pilyugin  \altaffilmark{1,2,4},
       Andreas Just \altaffilmark{2}
       }

\altaffiltext{1}{Main Astronomical Observatory, National Academy of Sciences of Ukraine, 
27 Akademika Zabolotnoho St., 03680 Kyiv, Ukraine}
\altaffiltext{2}{Astronomisches Rechen-Institut, Zentrum f\"ur Astronomie der Universit\"at Heidelberg, 
M\"onchhofstra{\ss}e 12--14, 69120 Heidelberg, Germany}
\altaffiltext{3}{National Astronomical Observatories of China, Chinese Academy of Sciences, 
20A Datun Rd., Chaoyang District, 1000128 Beijing, China}
\altaffiltext{4}{Kazan Federal University, 18 Kremlyovskaya St., 420008, Kazan, Russian Federation}

\shorttitle{On the influence of minor mergers on the radial abundance gradient in the disks of the Milky Way-like galaxies}
\shortauthors{Zinchenko et al.}

\begin{abstract}

We investigate the influence of stellar migration caused by minor
mergers (mass ratio from 1:70 to 1:8) on the radial distribution of
chemical abundances in the disks of Milky Way-like galaxies during
the last four Gyr.  A GPU-based pure N-body tree-code model without
hydrodynamics and star formation was used.  We computed a large set of
mergers with different initial satellite masses, positions, and orbital
velocities. We find that there is no significant metallicity change at
any radius of the primary galaxy in the case of accretion of a
low-mass satellite of 10$^9$~M$_{\odot}$ (mass ratio 1:70) except for
the special case of prograde satellite motion in the disk plane of the
host galaxy.  The accretion of a satellite of a mass
$\gtrsim3\times10^9$~M$_{\odot}$ (mass ratio 1:23) results in an
appreciable increase of the chemical abundances at galactocentric
distances larger than $\sim10$~kpc. The radial abundance gradient
flattens in the range of galactocentric distances from 5 to 15~kpc in
the case of a merger with a satellite with a mass
$\gtrsim3\times10^9$~M$_{\odot}$.  There is no significant change in
the abundance gradient slope in the outer disk (from $\sim15$~kpc  up
to 25~kpc) in any merger while the scatter in metallicities at a given
radius significantly increases for most of the satellite's initial
masses/positions compared to the case of an isolated galaxy.
This argues against attributing the break (flattening) of the
abundance gradient near the optical radius observed in the extended
disks of Milky Way-like galaxies only to merger-induced stellar
migration.

\end{abstract}


\section{Introduction}

It is well known that the heavy element abundance in the disks of the
Galaxy and other spiral galaxies decreases with increasing
galactocentric distance. This radial metallicity gradient was revealed
about 50 years ago through measurements of the metallicity in stars
and \ion{H}{2} regions \citep[e.g.,][]{Mayor1976, Searle1971,
Peimbert1979, Shields1978}.  The oxygen abundance gradient in nearby
spiral galaxies traced by \ion{H}{2} regions ranges from
0~dex~kpc$^{-1}$ to $\sim -0.1$~dex~kpc$^{-1}$
\citep[e.g.,][]{Vila-Costas1992, Zaritsky1994, Sanchez2013,
Pilyugin2014}.  Estimations of the iron and oxygen abundance gradient
in the Galaxy based on Cepheids and open clusters result in a slope of
approximately $-0.05$~dex~kpc${^{-1}}$ \citep{Andrievsky2002a,
Andrievsky2002b, Bragaglia2008, Lemasle2008, Pedicelli2009, Yong2012}.
However, the origin of the gradient and its relation to other
macroscopic parameters of galaxies are still open for discussion. 

There are indications that radial metallicity profiles in the Milky
Way and some giant spiral galaxies show a break within the optical
radius $R_{25}$, which corresponds the isophotal radius at a B-band
surface brightness of 25~mag~arcsec$^{-2}$. 
The iron abundance profile in the Galaxy traced by open clusters is
found to become shallower beyond 10--14~kpc by \citet{Bragaglia2008}
and \citet{Yong2012}. The earlier data on Cepheids confirm the
flattening of the iron gradient in the outer Galactic disk
\citep{Andrievsky2002a, Andrievsky2002b, Lemasle2008}, while recent
measurements of an extended sample of Cepheids show less evident (if
any) flattening of the gradient in the outer Galactic disk
\citep{Luck2011, Lemasle2013}.  OB star observations also do not
indicate an existence of a metallicity profile break in the outer
Galactic disk \citep{Daflon2004}.  The radial distribution of gas
phase oxygen and nitrogen abundances in the optical disks of the
majority of spiral galaxies traced by \ion{H}{2} regions can be well
fitted by a single exponential profile (linear in the term of
lg(X/H)) \citep{Pilyugin2014} while
possible profile slope changes are reported for some barred galaxies
\citep{Martin1995,Zahid2011,Scarano2011}.

Far-ultraviolet and near-ultraviolet observations by the {\sl Galaxy
Evolution Explorer (GALEX)} satellite reveal that a large fraction of
late-type galaxies in the local universe show significant star
formation beyond their optical radii $R_{25}$ \citep{GildePaz2007,Thilker2007}.
Those galaxies are referred to as extended ultraviolet disk (XUV-disk)
galaxies. Spectra of a number of \ion{H}{2} regions in the extended
disks of five spiral galaxies were measured recently 
revealing flat gradients in the extended disks
\citep{Bresolin2009,Goddard2011,Bresolin2012,Patterson2012,Werk2011}. 
\citet{Pilyugin2012} found the same radial gradient breaks for
nitrogen abundances in those galaxies and noted that the change in the
gradient slope is more distinct in the radial distribution of nitrogen
than of oxygen abundances.  Observations of OB stars in NGC~3621 also
support the metallicity profile flattening in the XUV-disk obtained
from the study of \ion{H}{2} regions \citep{Kudritzki2014}.
Thus, the abundance gradient is flatter in  the extended disks ($R > R_{25}$)
as opposed to the optical disks ($R < R_{25}$) in the case of a number of spiral
galaxies.

\citet{Bresolin2012} have advocated that it is unlikely that {\it in
situ} star formation could have produced and distributed enough heavy
elements to enrich the interstellar medium in the extended disks to
the presently observed values. \citet{Werk2010} have considered three
plausible mechanisms that may explain the relatively high metallicity
of the extended disks: radial redistribution of centrally generated
metals, strong galactic winds with subsequent fallback, and a past
interaction leading to the accretion of enriched gas. The last
mechanism, however, cannot be the sole explanation \citep{Werk2011}. 

The transport of the heavy elements from the inner to the extended
disks is also suggested as a possible explanation of flat radial
abundance gradients \citep{Werk2011, Bresolin2012}. 
Particularly, it was shown that the presence of a central bar and
spiral structure may lead to rather strong radial migration of stars
and gas in galactic disks
\citep{Friedli1994,Sellwood2002,Roskar2008,Schonrich2009,Minchev2010,Minchev2013,Minchev2014}
and a strong flattening during galaxy evolution in the [Fe/H] radial
profiles for the old stellar populations \citep{Minchev2014}. 
But while such a mechanism is quite effective in the inner disk, radial
migration studies that are in line with or going back to the analytic
prescription for radial migration strengths of \citet{Schonrich2009}
suggest very low migration for the outer disk in quiescent galaxies.

Radial migration can be also caused by the interaction and merging
with other galaxies.  Cosmological simulations demonstrate that minor
mergers should be a frequent and common process during the lifetime of
galaxies, in particular since z~$\sim$~1, and the number of satellites
that fall on a Milky Way-like galaxy strongly increases with
decreasing satellite mass \citep{Kazantzidis2008}. 
At the present day,  we observe the strong interaction of the Milky
Way with the Sagittarius dwarf galaxy \citep{Ibata1994, Ibata1995}.
The center of this dwarf spheroidal (dSph) galaxy with a mass of $\leq
10^9$~M$_{\odot}$ is located nearly behind our Galactic center at a
heliocentric distance of $\sim 26$~kpc \citep{Monaco2004}.  A
satellite with the mass of a few times 10$^{9}$~M$_{\odot}$ can
significantly heat the outer Galactic disk, excite the central spiral
structure and create a warp that can induce significant streams in
the velocity distribution \citep{SB2005,Quillen2009}.
\citet{Bird2012} considered the dynamical evolution of a Milky
Way-like galaxy with an accretion history motivated by cosmological
simulations and concluded that satellite perturbations appear to be a
distinct mechanism for inducing radial migration. Thus one may expect
a considerable role of minor mergers in the chemo-dynamical evolution
of galaxies during the last few Gyr.

So far the attention has been mainly focused on the behavior of
the radial metallicity gradient during the last few Gyr under the
``major merger'' scenario, i.e., interactions between massive galaxies
\citep{Kewley2010,Montuori2010,Rupke2010,Perez2011}, while less
attention has been paid to the possible influence of stellar migration
caused by a minor merger on the radial distribution of the metallicity
in the disks of spiral galaxies. 
In this study we will carry out high resolution N-body/TREE-GPU
simulations aimed at  examining the influence of pure stellar
migration caused by minor mergers (with a mass ratio of the mergers
from 1:8 to 1:70) on the radial distribution of metallicity in the
disks of large spiral galaxies out to 25~kpc. In particular, we will
investigate the role of the initial orbit orientation and of the mass
of the merged satellite in the formation of present-day metallicity
gradients.  We will explore whether satellite accretion can
plausibly be responsible for the flattening of the metallicity
gradient in Milky Way-like galaxy disks.

\section{Model and Code Description}

Our aim was to develop a simple model that can tell us how stellar
migration caused by mergers may have affected the radial distribution
of metallicity in the disks of a Milky Way-like galaxy.  In order to
see clearly the purely dynamical mixing effect (migration) and to
reduce the number of free parameters in our model we adopt a number of
simplifications: 
(1) our merger system is purely based on N-body simulations, without gas; 
(2) we do not assume star formation; 
(3) the initial dynamic and kinematic parameters of a Milky Way-like 
galaxy are assumed to be close to the Milky Way parameters at the 
present time; 
(4) the initial metallicity distribution of our model is based on 
the measured abundance distributions in spiral galaxies traced by 
\ion{H}{2} regions and young stellar populations; and
(5) we neglect the possible dilution of disk metallicity by satellite 
stars deposited in the disk of the host galaxy. In this sense the 
satellite is considered just as a dark matter sub-halo.

We consider galaxy evolution only during the last four Gyr since the
initial conditions of our model are close to those at the present
time.  Cosmological simulations of the formation of a Milky Way-like
galaxy support our assumption that the disk of a Milky Way-like galaxy
remained quite stable during the last few Gyr \citep{Bird2013}.
This approach leads us to a model with a strongly reduced number of
degrees of freedom for the investigation of the influence of the
migration of collisionless particles caused by minor mergers. 

For our Milky Way-like galaxy we adopt the \citet{KD1995} ``MW-D'' model
of a disk galaxy with some slightly changed parameters. In our model
we represent the galaxy with 2 million disk particles with a total
mass of 5.75 $\times$ 10$^{10}$ M$_{\odot}$, 0.4 million bulge
particles with a total mass of 1.29 $\times$ 10$^{10}$ M$_{\odot}$ and
5 million of dark matter halo particles with a total mass of 9.89
$\times$ 10$^{11}$~M$_{\odot}$. We take the exponential scale length
of the disk to be $R_d$ = 2.97~kpc, the disk scale height $z_d$ =
0.225~kpc, and the central radial velocity dispersion $\sigma_r$ =
123~km~s$^{-1}$. With this choice of parameters the stellar surface
density at the solar vicinity is $\sim$57~M$_{\odot}$pc$^{-2}$ and
Toomre's stability parameter $Q$ \citep{Toomre1964} is 2.6.  We report
the total list of our host galaxy model parameters in 
Table~\ref{tableMW} and Table~\ref{tableMW-2}. 

\begin{table}[!ht]
\begin{center}
\caption{Galaxy Model Parameters.}
\label{tableMW}
\begin{tabular}{l|r}
\tableline
Bulge cut-off potential, $\Psi_{\rm c}$    		& -4.7		\\
Bulge velocity dispersion, $\sigma_{\rm b}$ 	&  1.29  	\\
Bulge central density, $\rho_{\rm b}$       	&  125 		\\
\hline
Halo central potential, $\Psi_0$      				& -6.7  	\\
Halo velocity dispersion, $\sigma_0$  				&  1.45  	\\
Halo potential flattening, $q$        				&  1.0  	\\
Halo concentration, $C$               					&  0.1 		\\
Characteristic halo radius,  $R_{\rm a}$    	&  0.7  	\\   
\hline
Disk mass, $M_{\rm d}$                      	&  1.1		\\ 
Disk scale radius, $R_{\rm d}$              	&  0.66		\\
Disk truncation radius, $R_{\rm out}$         &  7.0		\\
Disk truncation width, $\delta R_{\rm out}$  &  0.5		\\
Disk scale height, $z_{\rm d}$              	&  0.05		\\
\tableline\end{tabular}
\tablecomments{All values of the dimensionless model were adopted \\ from \citet{KD1995}. 
The units of length, velocity, \\ and mass used here are: 
{\it RU} = 4.5~kpc, {\it VU} = 220~km~s$^{-1}$,\\
{\it MU} = 5.1 $\times$ 10$^{10}$ M$_{\odot}$, respectively.}
\end{center}
\end{table}

\begin{table}[ht]
\begin{center}
\caption{Galaxy Component Parameters.}
\label{tableMW-2}
\begin{tabular}{l|r}
\tableline
Number of bulge particles, $N_{\rm b}$         	&  4$\times$10$^5$ \\
Bulge mass, $M_{\rm b}$                      	&  1.29$\times$10$^{10}$~M$_{\odot}$		\\ 
Bulge particle mass, $m_{\rm b}$               	&  3.23$\times$10$^4$~M$_{\odot}$		\\
\hline
Number of halo particles, $N_{\rm h}$         	&  5$\times$10$^6$ \\
Halo mass, $M_{\rm h}$                      	&  9.89$\times$10$^{11}$~M$_{\odot}$		\\ 
Halo particle mass, $m_{\rm h}$               	&  19.78$\times$10$^4$~M$_{\odot}$		\\
\hline
Number of disk particles, $N_{\rm d}$          	&  2$\times$10$^6$ \\
Disk mass, $M_{\rm d}$                      	&  5.75$\times$10$^{10}$~M$_{\odot}$		\\ 
Disk particle mass, $m_{\rm d}$                	&  2.88$\times$10$^4$~M$_{\odot}$		\\
Disk scale radius, $R_{\rm d}$              	&  2.97 kpc		\\
Disk scale height, $z_{\rm d}$              	&  225 pc		\\
\tableline\end{tabular}
\end{center}
\end{table}

The initial distribution of the particles in the satellite galaxy
(``perturber'') is given by a simple spherically symmetric Plummer
model with $N_{sat}=33000-300000$ particles and with an initial total
mass from 10$^{9}$~M$_{\odot}$ ($N_{sat}=33000$) to $9 \times
10^{9}$~M$_{\odot}$ ($N_{sat}=300000$).  The initial half-mass radius
was chosen to correspond to one third of the power of the
initial mass of the satellite from $R_{\rm hm}$ = 0.15~kpc to
0.31~kpc. In this way we keep the initial ratio of the tidal radius to
the half-mass radius of the satellite galaxy roughly constant (in our
case it was $\approx$~25). 

Oxygen was chosen as ``metallicity indicator''. We adopted as an
initial condition that  all the particles at a given galactocentric
distance have the same metallicity, which was taken from the radial
abundance gradient relation.  We assume the central oxygen abundance
to be 12+lg(O/H) = 8.8 and the slope of the gradient to be
$-0.05$~dex~kpc$^{-1}$, i.e.,
\begin{equation}
12 + \lg(\frac{\rm O}{\rm H}) = 8.8 - 0.05 \cdot  R 
\end{equation}
where $R$ is galactocentric distance in kpc for the disk particles.
These values are based on the measured abundance distributions in
spiral galaxies \citep[e.g.,][]{Vila-Costas1993, Zaritsky1994,
Maciel2010, Pilyugin2003, Pilyugin2004, Pilyugin2014, Sanchez2013}. 
This simplification seems to be justified since detailed studies of
the age-metallicity relation of stars in the Galactic disk (solar
vicinity) show an almost constant metallicity for the last four Gyr,
accompanied by substantial scatter (e.g., \citet{Haywood2013}, their
figure~9; \citet{Casagrande2011}, their figure~18). 

Thus, on the one hand, our particles can be considered as ``stars''
since an N-body simulation is used and the metallicity of the particle
does not change during the evolution.  On the other hand, they can be
considered as ``gas clouds'' since the metallicities of all the
particles at a given radius at the initial time  are taken to be the
same.  It should be noted that since our simulations deal with a pure
N-body problem without involving any chemical enrichment, our results
can be applied to other elements as well, not only to oxygen. There is
only one restriction: the process of chemical radial mixing must be
done by objects that can represented as a ``N-body particles''.

\subsection{Code description}

We ran all simulations using our version of the N-body TREE-GRAPE
(GRAvity PipEline) code \citep[][section 4.1]{FMK2005}. We also ported
the original GRAPE-based tree code to many different hardware
platforms including the CPU (with multi-thread usage under the SSE or
AVX instructions) and also to the recent NVIDIA Graphics Processing
Unit (GPU) platform using the Compute Unified Device Architecture
(CUDA).  In order to efficiently use the current acceleration hardware
(GPU), we employed a modified tree algorithm, originally developed by
\citet{BH1986}, which was first used on GRAPE-1A by \citet{M1991} and
\citet{FIM1991}.

On a typical desktop hardware (CPU: i5-2500K with 4 cores @ 3.3 GHz +
GPU: GeForce GTX 570 with 480 cores @ 1.46 GHz) we get the results for
the full self-gravity force calculation routine with the typical
tree-construction parameters ($\theta$ = 0.5, $n_g$ = 3500,
\citep[][section 4.1.1]{FMK2005}) for $N = 1$~M particles with the
initial Plummer distribution in $\approx$ 2 sec. Our speed is quite
comparable to the most advanced and recent fully GPU tree code
implementation ({\tt bonsai2}: \cite{BGPZ2012a, BGPZ2012b}), which does
one full-force calculation on the same hardware for the same
particle distribution and with the same opening angle in $\approx$ 1
sec. In the last years we already used and extensively tested our
hardware-accelerator-based gravity calculation routine in a few of our
galactic dynamics projects and get quite accurate results with a
good performance \citep{SB2005, BJBB2008, PGBSD2010, PGBCS2011}.

The current set of simulations was carried out with the GPU version of
the code using local GPU clusters available at the authors'
institutions (ARI: {\tt kepler}, MAO: {\tt golowood}, NAOC: {\tt
laohu}). 

We call the current version of our N-body code {\tt ber-gal0}. The
source code of the program is publicly available from the FTP
site\footnote{\tt ftp://ftp.mao.kiev.ua/pub/users/berczik/ber-gal0/}
of one of the authors.  In our production runs for the gravitational
force calculation we use a very conservative opening angle $\theta$ =
0.4. A gravitational softening parameter $\epsilon$ = 40~pc was
adopted for all the particles. We use the simple leap-frog integration
scheme with a fixed time-step $\Delta t = 0.313$~Myr to advance the
particle positions and velocities during the calculation. One basic
run covering up to 4~Gyr with $\sim7.4$ million particles takes
$\sim55$ hours. The total number of models presented in the current
paper is 34, so the total computational time was $\sim$78 days.

\subsection{Disk stability}

\begin{figure*}[!Ht]                                                          
  \begin{center}                                                              
  \includegraphics[width=0.40\textwidth]{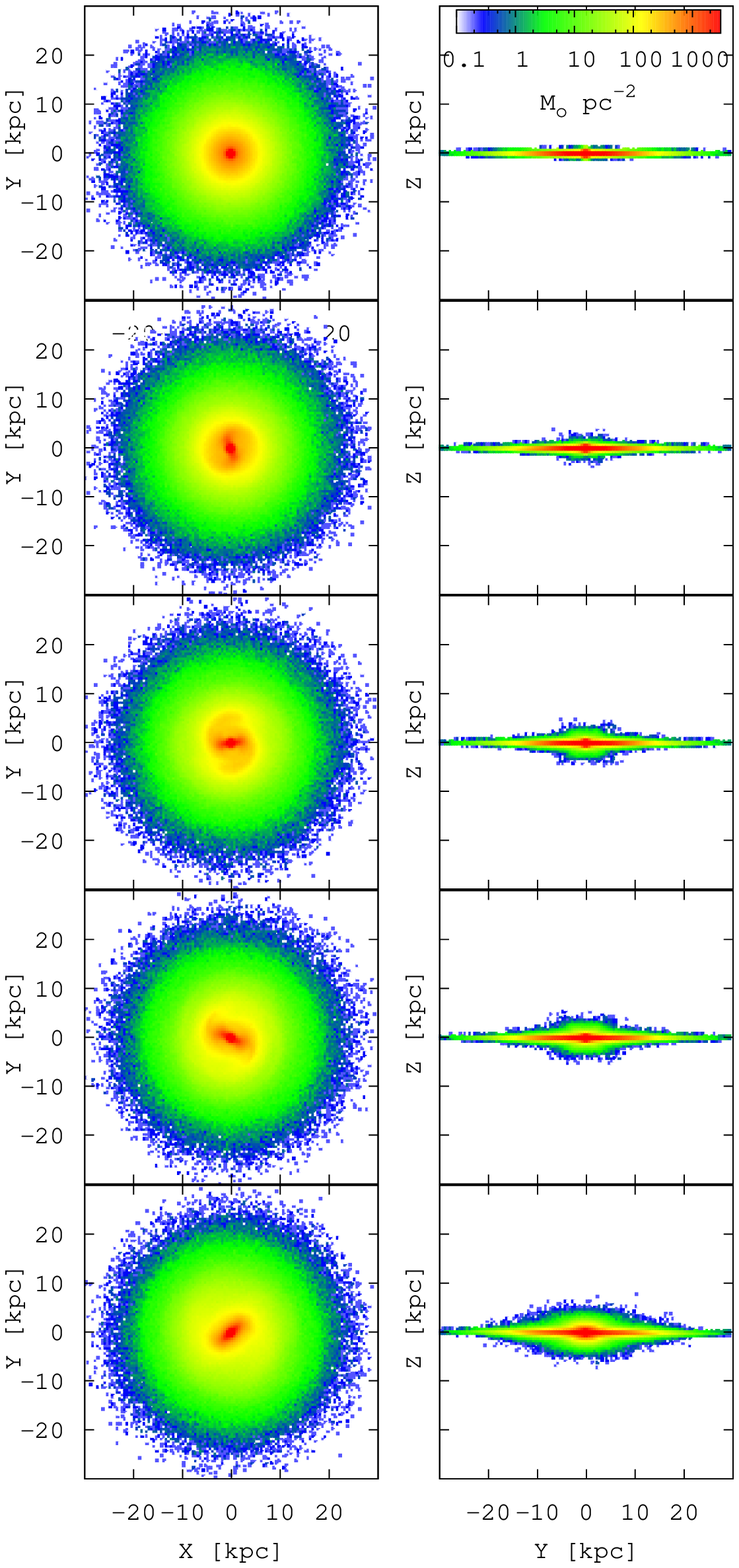} 
  \includegraphics[width=0.40\textwidth]{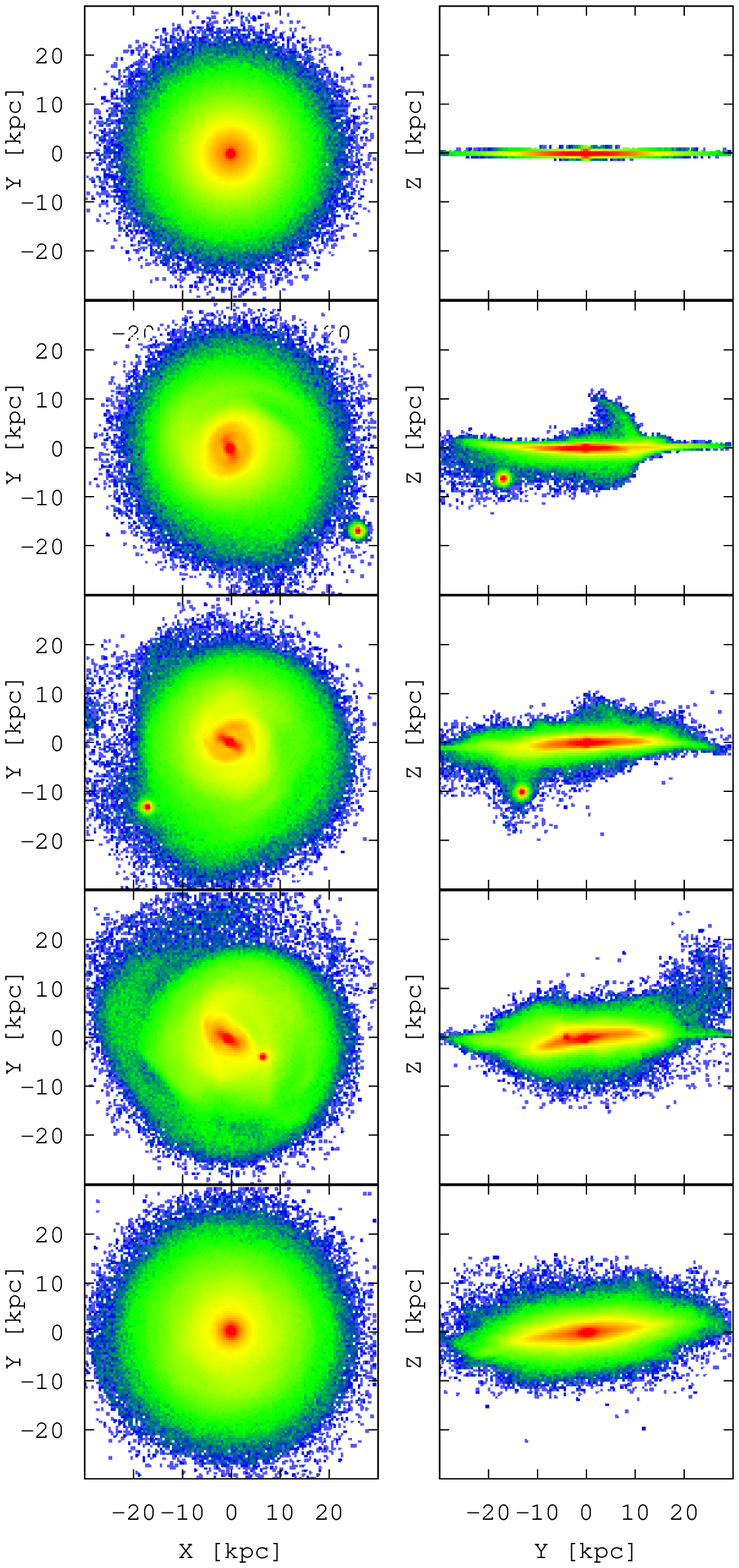} 
  \end{center}                                                                
  \caption{Comparison of the log-scale column density image for 
	disk+bulge+satellite components in the model A00 (1st and 2nd column) 
	and model C02 (3rd and 4th column). Top to bottom: density at 
	$t = 0$, 0.4, 0.8, 1.2, and 4~Gyr.} 
  \label{fig_image}                                                    
\end{figure*}

First of all, we computed the evolution of the host (Milky Way-like)
galaxy without a satellite (no ``perturber'') in order to test the
stability of the structure and the radial oxygen abundance
distribution during 4~Gyr of evolution.  The first and second columns
of Figure~\ref{fig_image} show the column-density evolution of the
disk+bulge components of the galaxy model at $t = 0$, 0.4, 0.8, 1.2, and
4~Gyr.  The face-on column density maps (left column of
Figure~\ref{fig_image}) show that the isolated galaxy forms a
long-lived bar. 

\begin{figure*}[htbp]
  \begin{center}
  \includegraphics[width=0.3\linewidth]{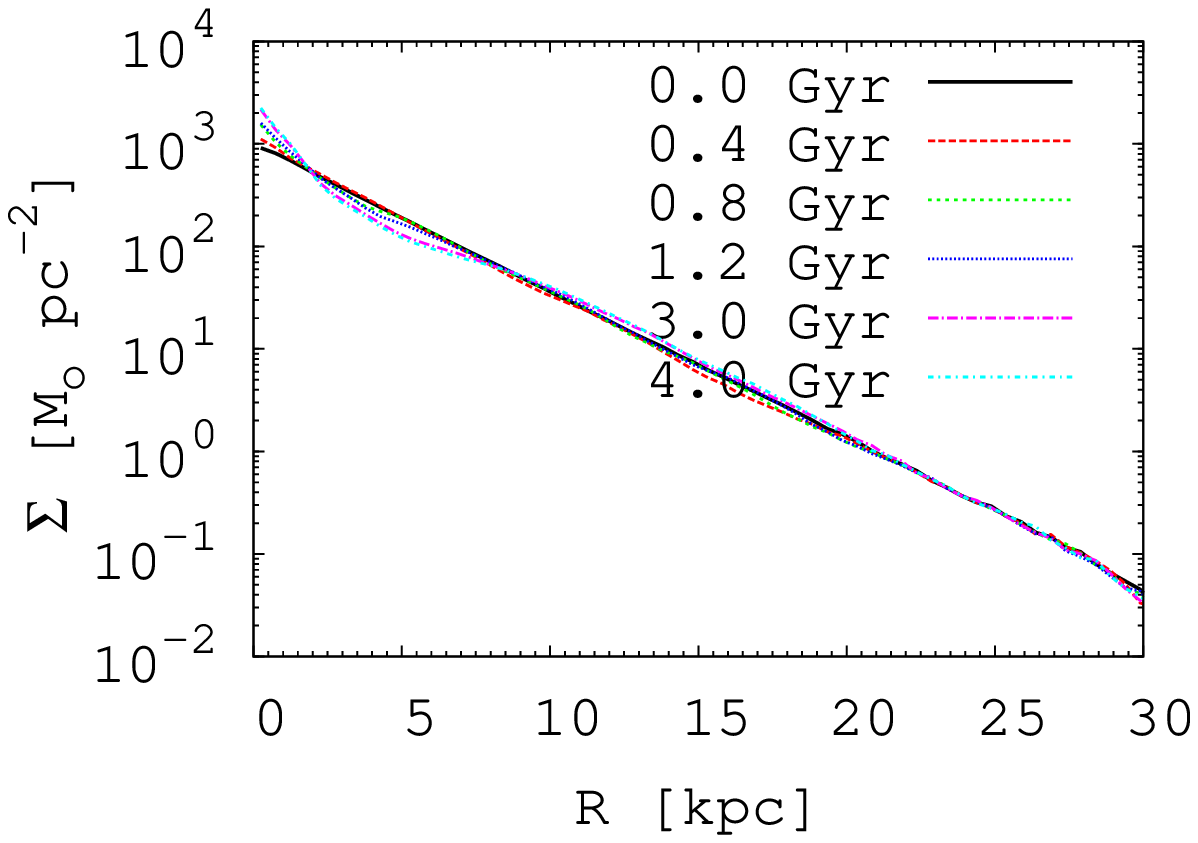}
  \includegraphics[width=0.3\linewidth]{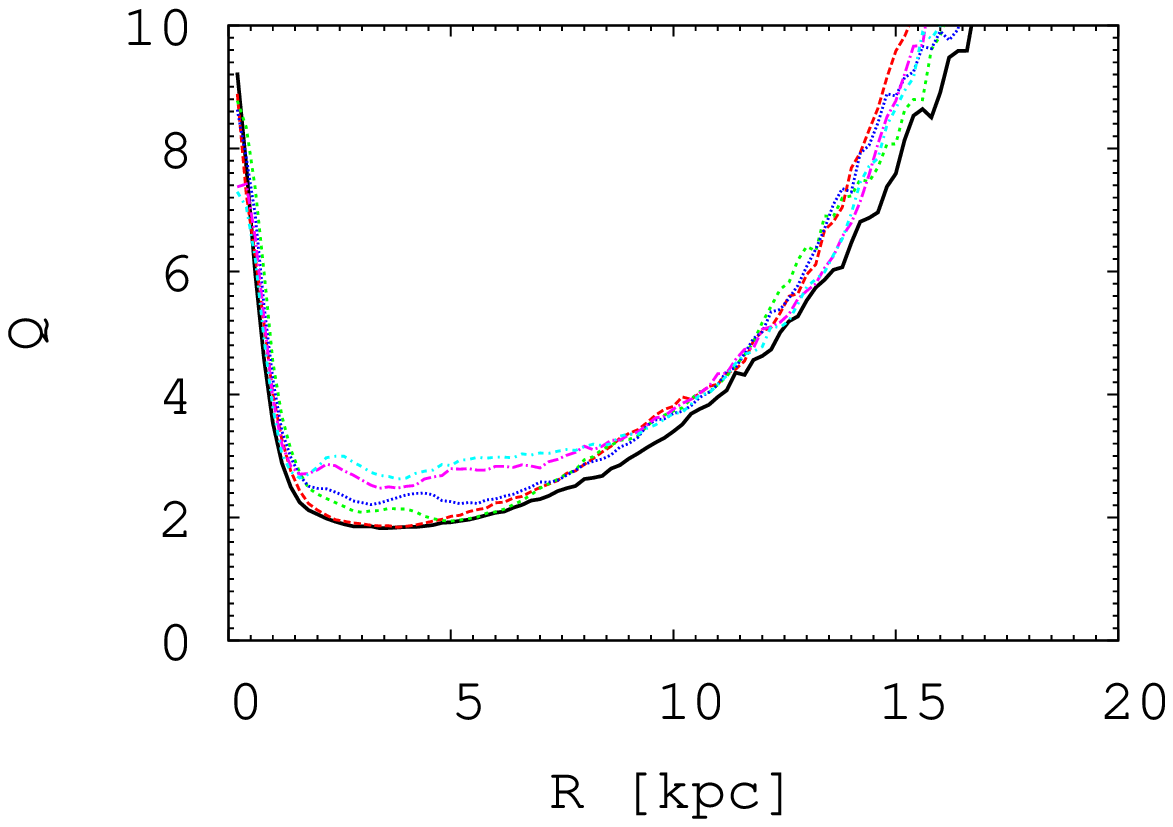}  
  \includegraphics[width=0.3\linewidth]{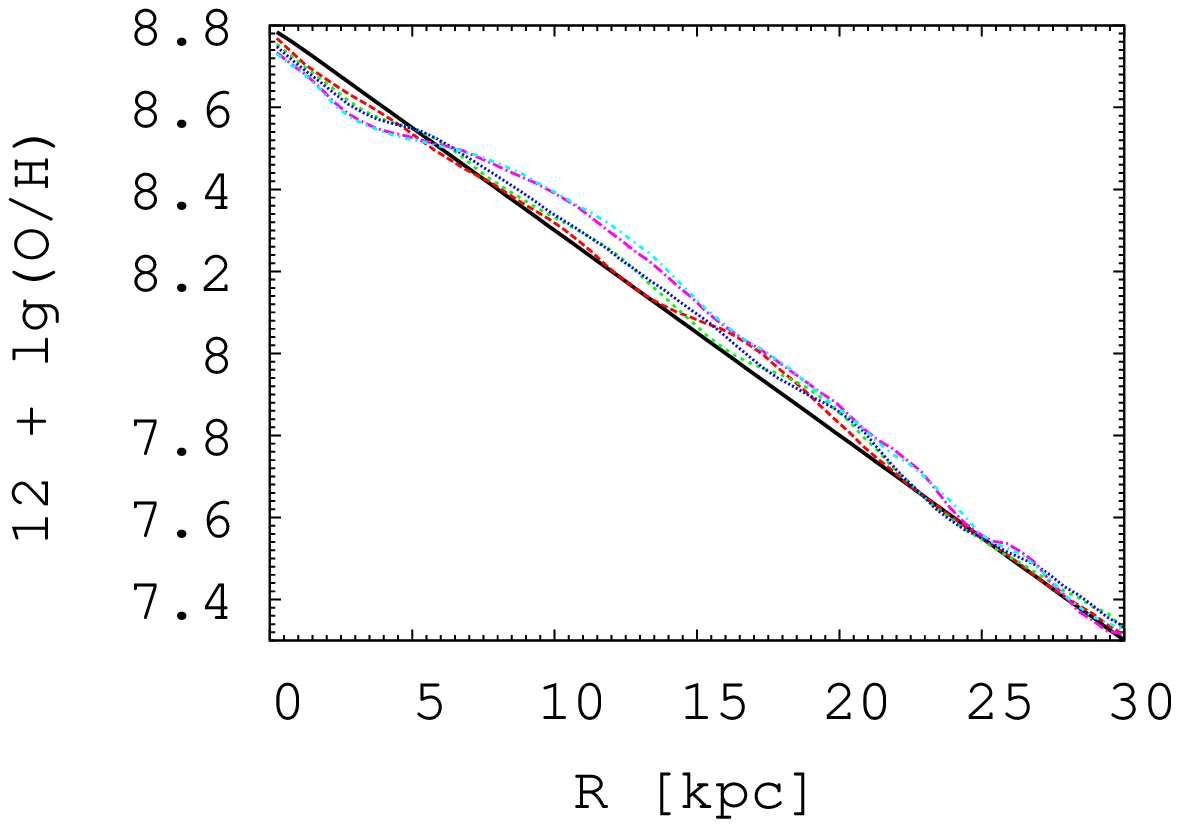}
  \includegraphics[width=0.3\linewidth]{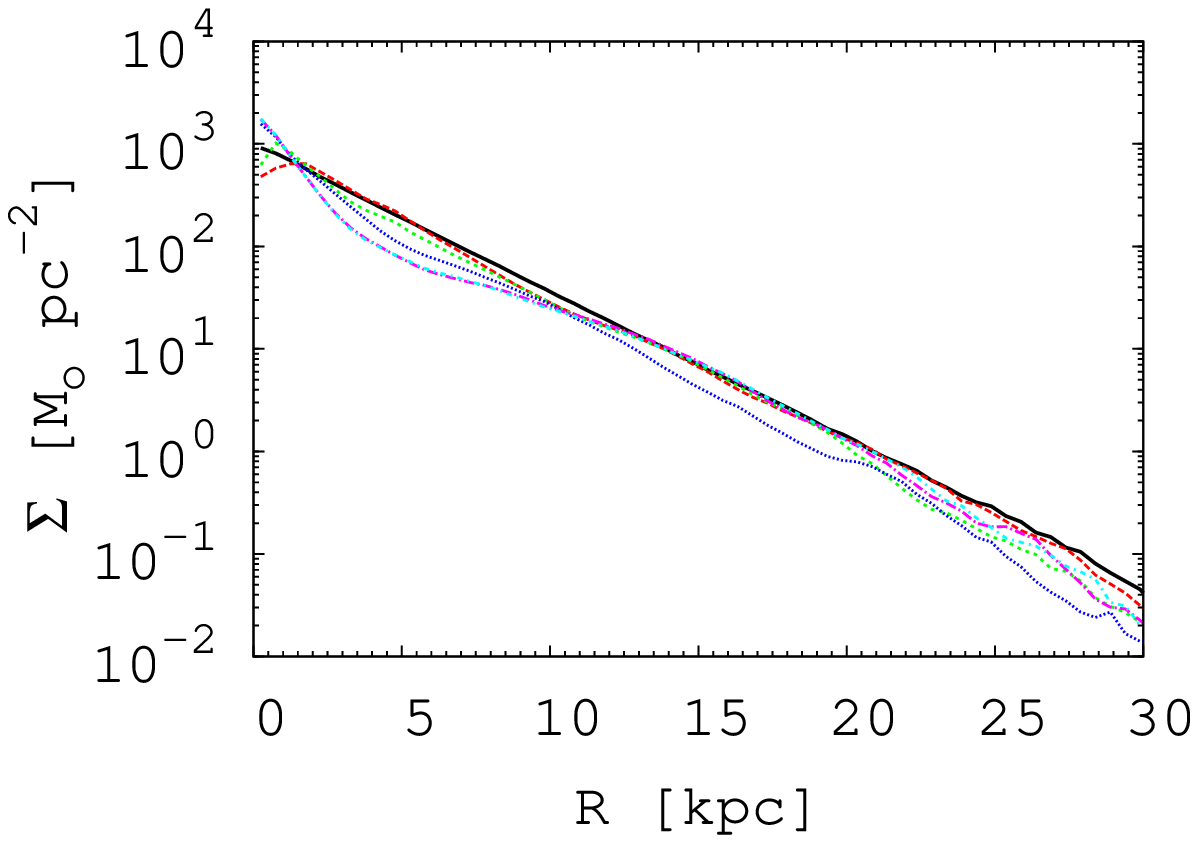}
  \includegraphics[width=0.3\linewidth]{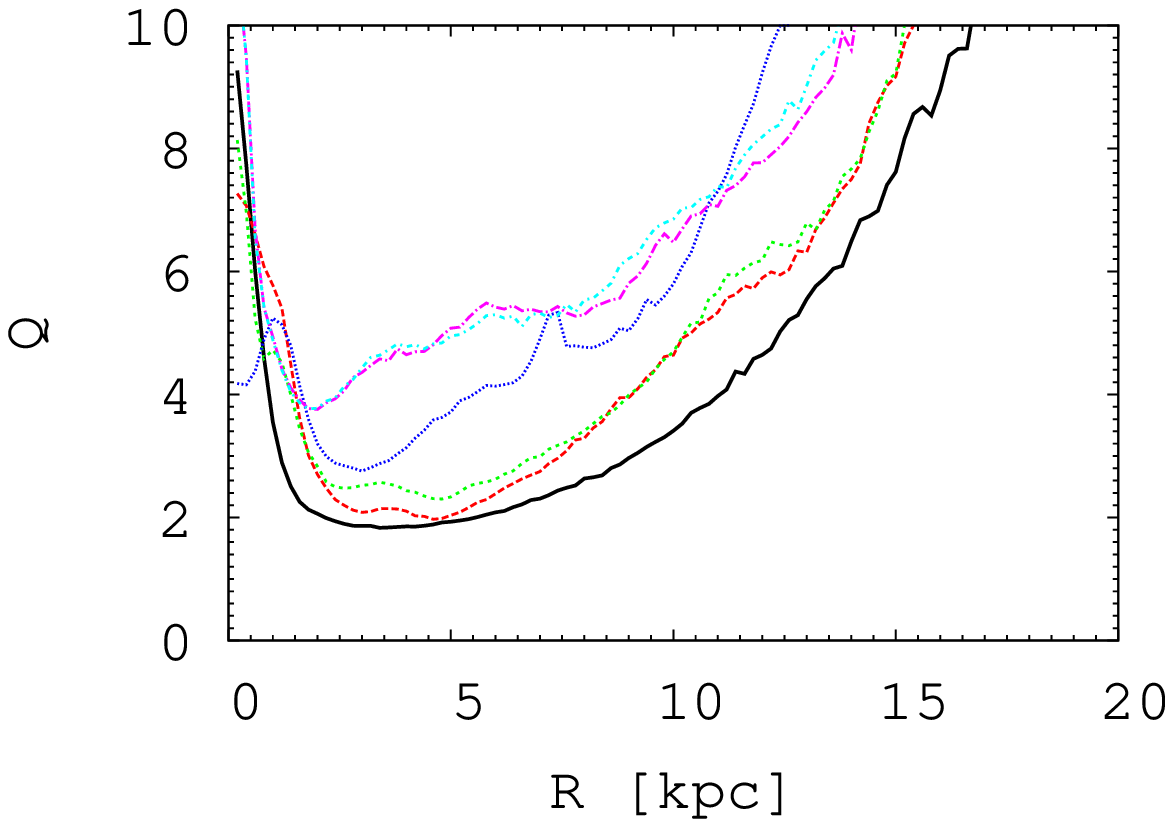}  
  \includegraphics[width=0.3\linewidth]{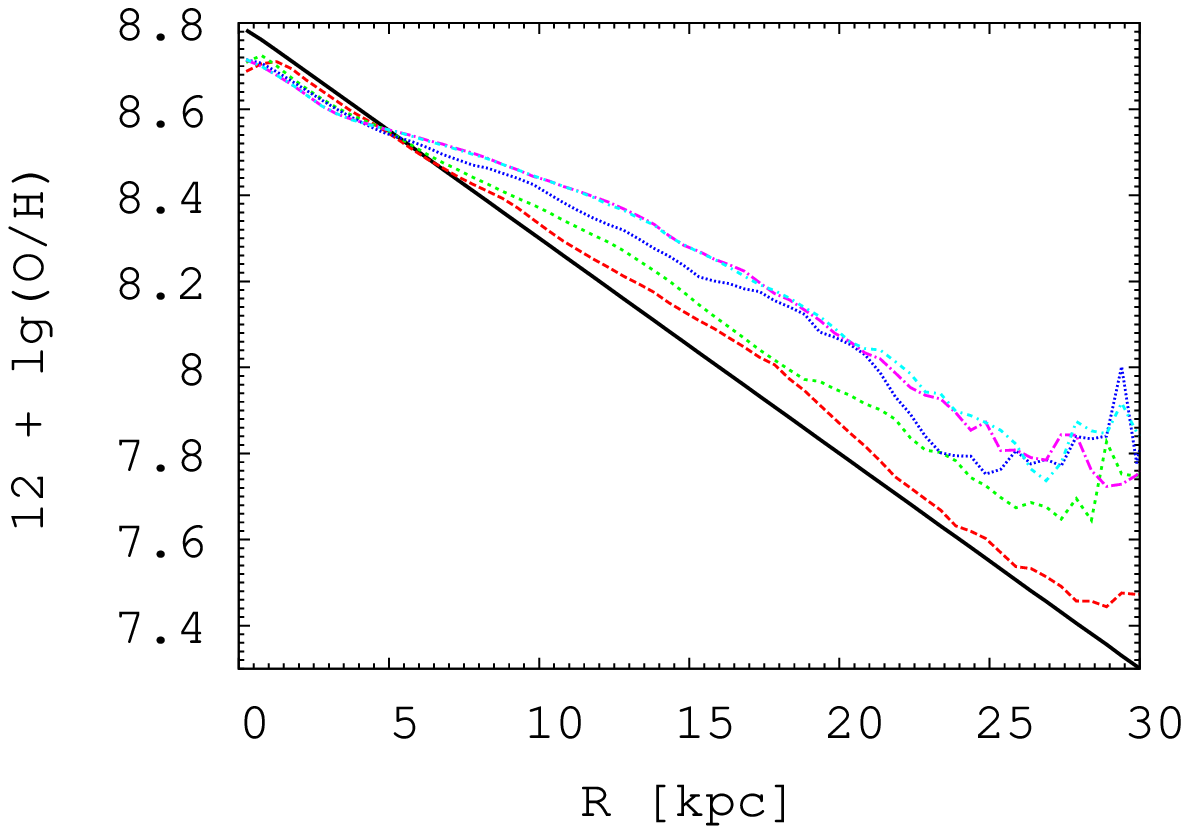}
  \end{center}
  \caption{Example of the evolution of the disk radial surface density (left column), 
        Toomre's stability parameter $Q$ (middle column)
        and oxygen abundance profiles (right column) during 
	the 4 Gyr considered in our simulations. Top: galaxy without satellite. 
	Bottom: galaxy with a merging satellite with a mass of 
	$6 \times 10^{9}$ M$_{\odot}$ (mass ratio 1:12). 
	The satellite's orbital inclination is $i = 30^\circ$ (model C02).}
  \label{fig_oh-grad-evol}
\end{figure*}

Changes in the disk's radial surface density, Toomre's stability
parameter $Q$, and in the oxygen abundance profiles during the 4~Gyr
of isolated host galaxy evolution are presented in the top panels of
Figure~\ref{fig_oh-grad-evol}.  The radial surface density profiles of
the disk show that the formation of a bar leads to radial mixing in
the inner part of galactic disk.  According to Toomre's stability
parameter $Q$, our disk model is stable against non-axisymmetric
perturbations (see Fourier analysis presented on Figure~\ref{fig_fft})
with a minimal value $Q = 1.9$ at a radius of 3.6~kpc for the initial
isolated disk setup with an increase of the minimal $Q$ value during
the run.

Nevertheless, the variations of the disk's radial surface density and
radial metallicity profile during 4~Gyr of evolution of an undisturbed
galaxy are small (top panels of Figure~\ref{fig_oh-grad-evol}).  The
maximum value of changes in the oxygen abundance across the entire
range of galactocentric distances also does not exceed $\sim 0.1$~dex
during 4~Gyr of evolution of our undisturbed galaxy. This can be
considered as evidence that the host galaxy model is sufficiently
stable and can be used to study changes in the radial abundance
distribution due to migration caused by the merging (interaction) with
a satellite. 

\subsection{Stability against disk and code parameters}

\begin{figure*}[htbp]
  \begin{center}
  \includegraphics[width=0.3\linewidth]{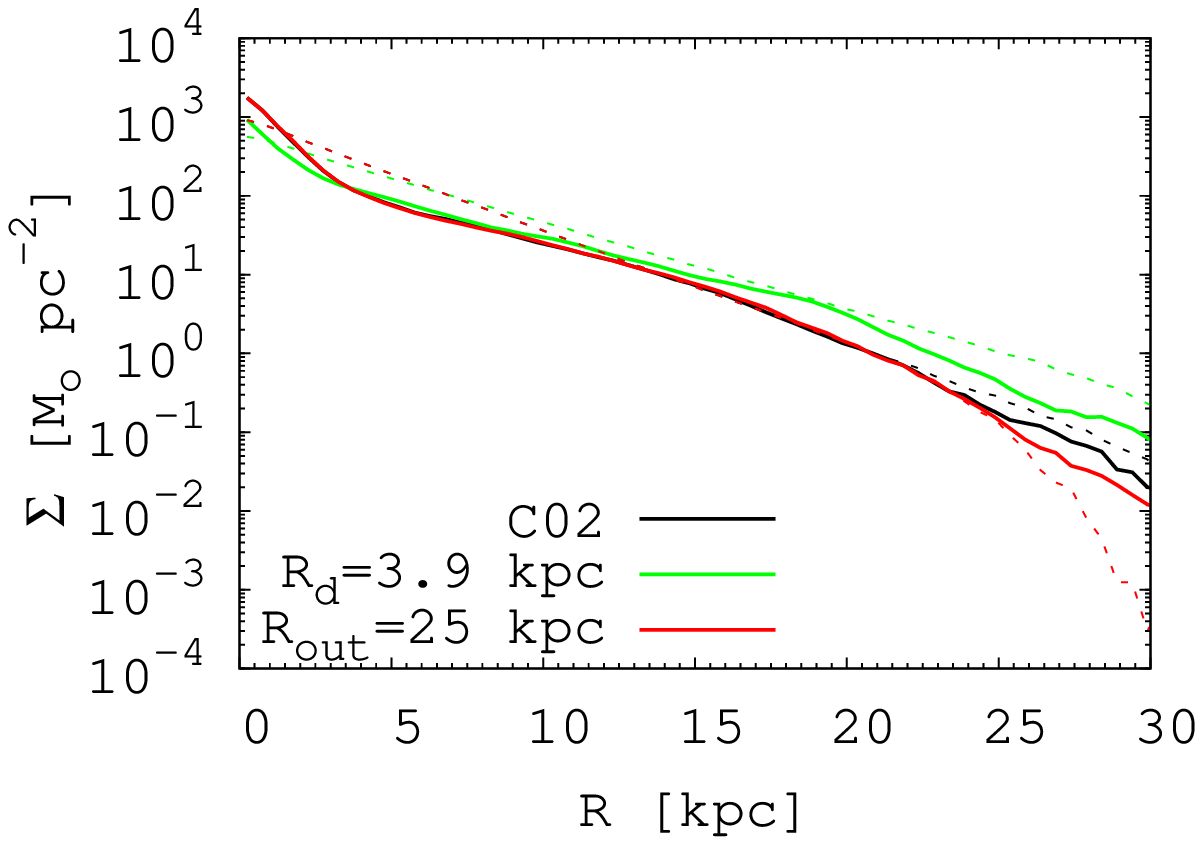} 
  \includegraphics[width=0.3\linewidth]{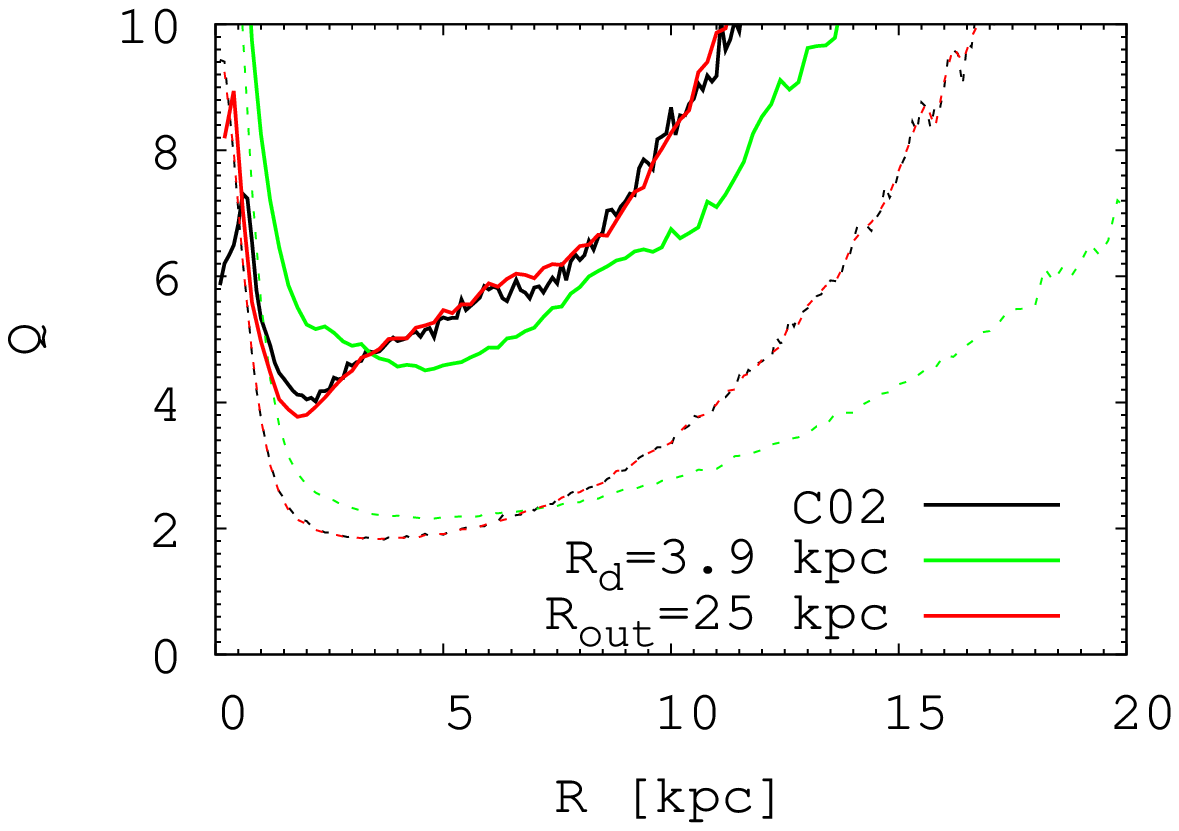} 
  \includegraphics[width=0.3\linewidth]{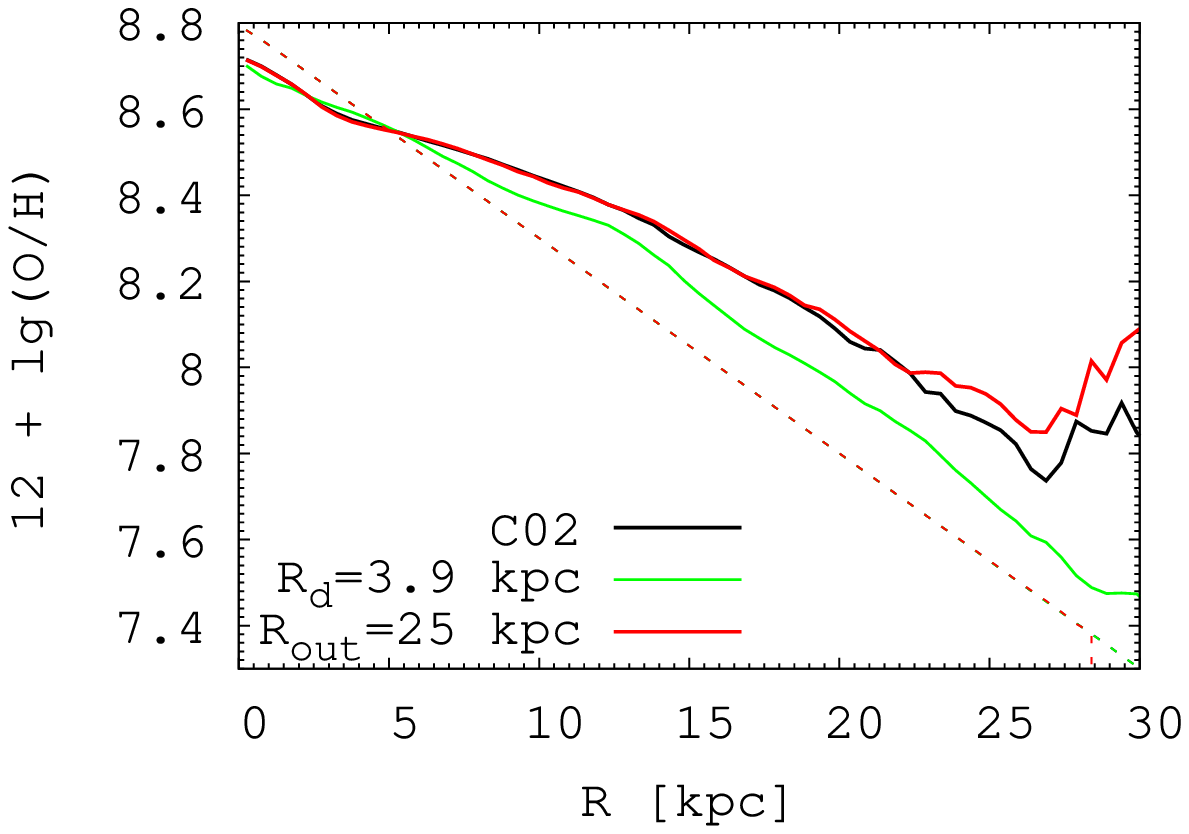} 
  \includegraphics[width=0.3\linewidth]{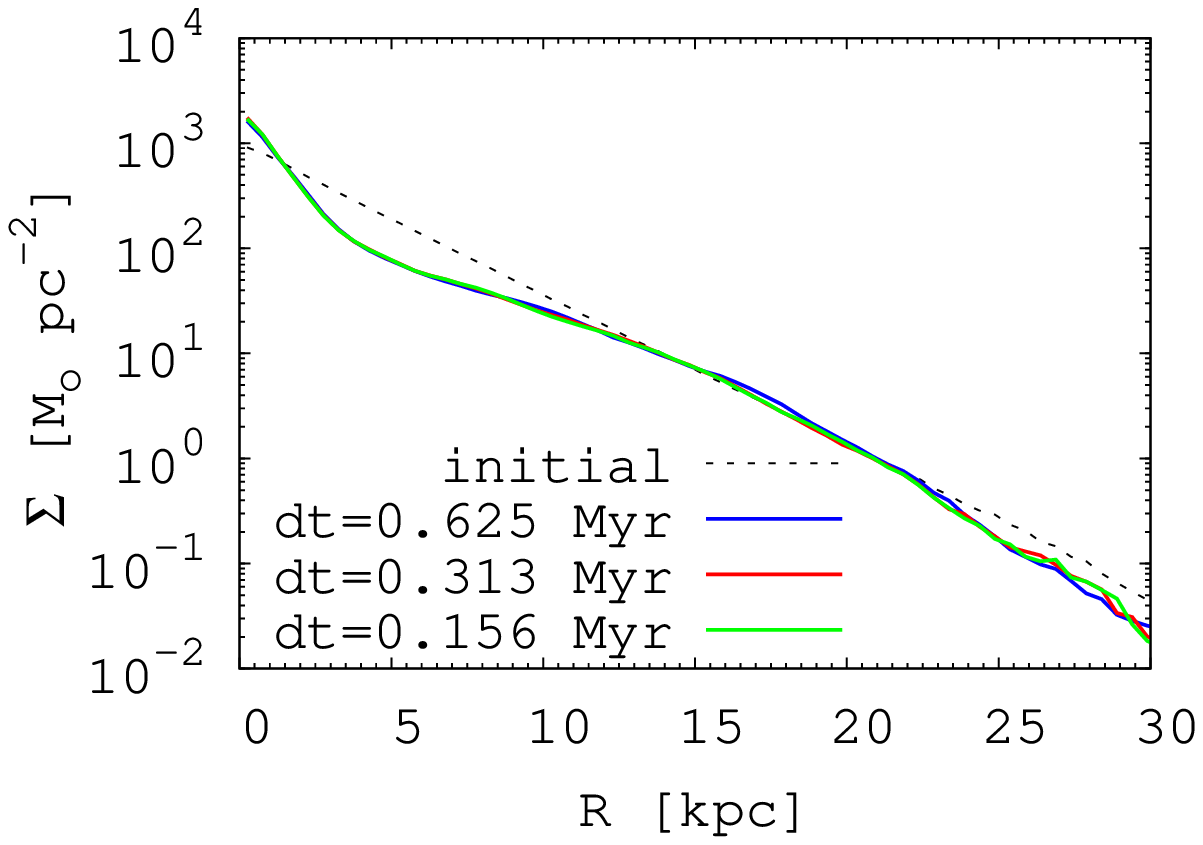} 
  \includegraphics[width=0.3\linewidth]{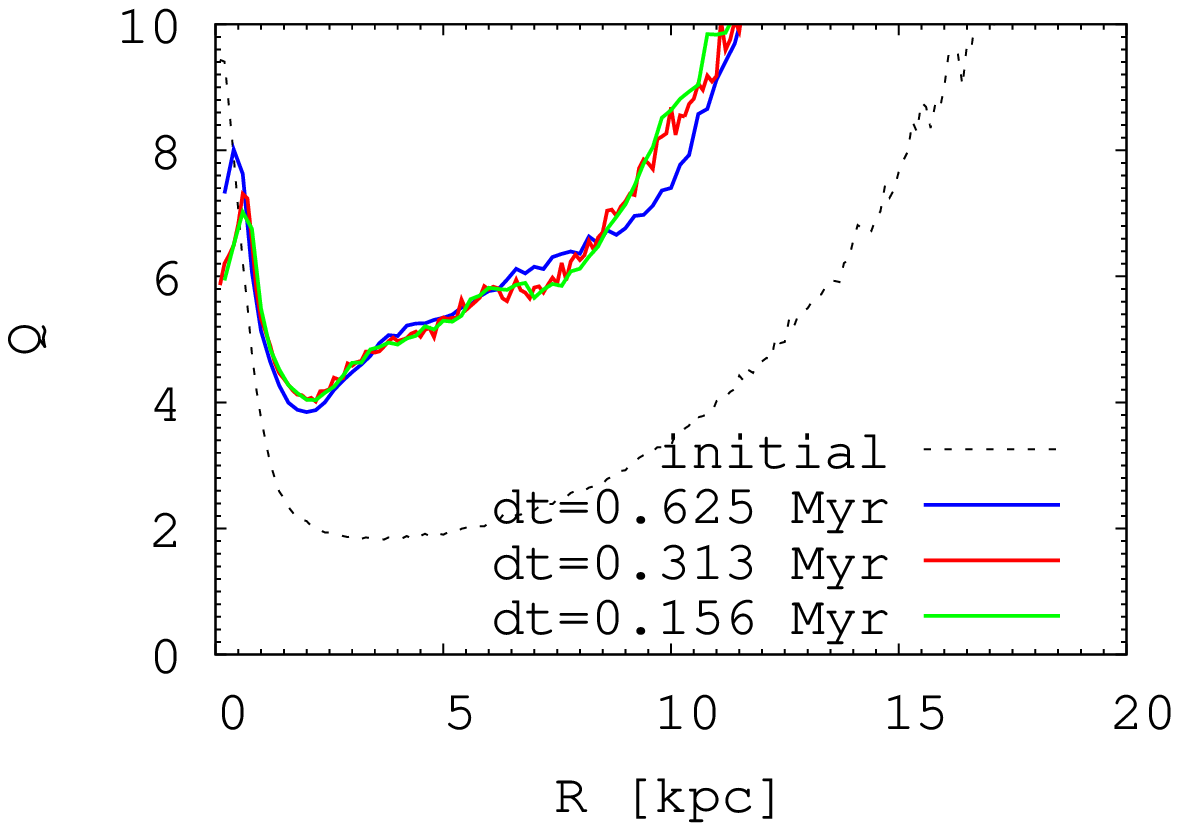} 
  \includegraphics[width=0.3\linewidth]{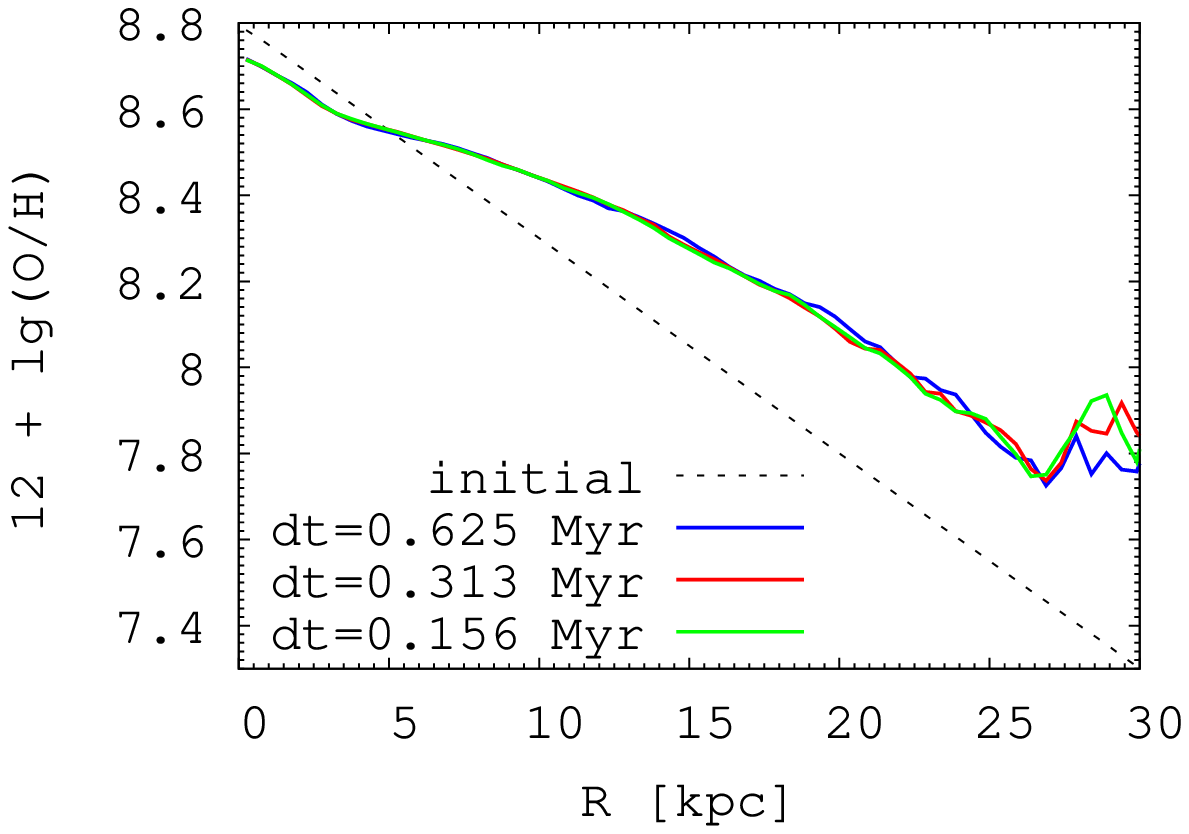} 
  \end{center}
  \caption{ Top: Influence of the variation of the parameters of the disk
on the disk's radial surface density (left column), 
        Toomre's stability parameter $Q$ (middle column), and oxygen abundance profiles (right column).
        Profiles for model C02 (a galaxy with a basic disk scale radius
$R_{\rm d}=3$~kpc and disk truncation radius $R_{\rm out} =31.5$~kpc) 
        (black lines), and models with $R_{\rm d}=3.9$~kpc (green lines)
and $R_{\rm out}=25$~kpc (red lines) are presented.
        The dotted lines indicate initial profiles. The solid lines
show profiles at $t=4$~Gyr.
        Bottom: Stability of the disk radial surface density (left column), 
        Toomre's stability parameter $Q$ (middle column),
        and oxygen abundance profiles (right column) for different
time steps.
        Profiles for model C02 with $\Delta t=0.625$~Myr (blue line),
$\Delta t=0.313$~Myr (red line), and $\Delta t=0.156$~Myr (green line) are presented.
        The dotted lines indicate initial profiles; solid lines
indicate profiles at $t=4$~Gyr.}
  \label{fig_oh-grad-evol-test}
\end{figure*}

The top panels of Figure~\ref{fig_oh-grad-evol-test} show the
influence of the variation of the disk parameters on the disk's radial
surface density (left column), Toomre's stability parameter $Q$
(middle column), and oxygen abundance profiles (right column).
Profiles for model C02 (a galaxy with a basic disk scale radius $R_d =
3$~kpc and disk truncation radius $R_{\rm out} =31.5$~kpc, which
merged with a satellite of a mass of $6\times 10^{9}$ M$_{\odot}$
(black lines), and models with $R_d =3.9$~kpc (green lines) and $R_{\rm
out}=25$~kpc (red lines) are presented. The dotted lines indicate the initial
profiles, and the solid lines show profiles at 4 Gyr of evolution.  
The initial slopes of the oxygen abundance gradient are in all cases 
--0.05~dex~kpc$^{-1}$.
It is evident that increasing the disk scale radius by 30$\%$ increases
also the surface density of the outer disk, which leads to a decrease
of Toomre's stability parameter $Q$ in the outer disk, but makes it
more stable against changes of the oxygen abundance induced by
mergers. 
This behavior can be explained as follows. The level of the
increase of the abundance in the outer disk depends on the 
ratio of the number of high-metallicity particles that migrated from the 
inner region of the disk to the number of particles in the outer disk. Since 
increasing the $R_d$ increases the number of non-migrated particles (surface density) 
in the outer disk, the increase the oxygen abundance in the outer 
disk is less for a disk with higher $R_d$.
A reduction of the disk size from 31.5~kpc to 25~kpc shows
no appreciable changes in the oxygen abundance profile within the
reduced disk radius, i.e., up to 25~kpc.

The bottom panels of Figure~\ref{fig_oh-grad-evol-test} show the
stability of the disk's radial surface density (left column), Toomre's
stability parameter $Q$ (middle column), and the oxygen abundance
profiles (right column) for different time steps in the simulation.
Profiles for model C02 with $\Delta t=0.625$~Myr (blue line), $\Delta
t=0.313$~Myr (red line), and $\Delta t=0.156$~Myr (green line) are
presented.  The dotted lines indicate initial profiles. The solid
lines indicate profiles at $t=4$~Gyr.

\section{Satellite initial orbits and velocity}

We ran four sets of simulations with satellite masses ranging from
10$^{9}$ M$_{\odot}$ to 9 $\times$ 10$^{9}$ M$_{\odot}$ (mass ratios
of 1:70 to 1:8). Each set consists of 7 runs with different initial
positions of the satellite galaxy. The initial distance from the
center of the host galaxy was taken to be 40~kpc 
for all the runs. The initial three-dimensional positions and velocity
directions were chosen in such a way that the satellite orbit
inclination is equal to the fixed angles $i = 0^\circ$, $30^\circ$,
$60^\circ$, $90^\circ$, $120^\circ$, $150^\circ$, and 180$^\circ$. The
inclinations $i>90^\circ$ correspond to a retrograde orbital motion of
the satellite. The initial absolute value of the satellite velocity
for our runs is $V = 100$~km~s$^{-1}$ (see discussion below). The
parameters of the simulations are summarized in Table~\ref{table2}.

To check how the orbital elements affect the radial abundance
distribution we ran a set of five simulations with an initial
satellite velocity from $V = 0$~km~s$^{-1}$ to 200~km~s$^{-1}$, with a
fixed initial distance from the galactic center of 40~kpc, with a
fixed initial orbital inclination $i = 30^\circ$, and with a fixed
mass of M$_{\rm sat} = 6 \times 10^{9}$~M$_{\odot}$.  The parameters
of this set of simulations are summarized in the Table~\ref{table1}.
The results are discussed below.

\begin{table*}[htb]
\small
\begin{center}
\caption{Model runs with different satellite initial velocities and the resulting disk oxygen abundance gradients.}
\label{table1}
\begin{tabular}{c|crcccccc}
\tableline
  Model &  Mass [$\times10^{9} M_{\odot}$]  &  $V$ [km~s$^{-1}$] & $\Delta$$\lg$(O/H)$_{\rm all}$ & $\Delta$$\lg$(O/H)$_{\rm inner}$ & $\Delta$$\lg$(O/H)$_{\rm outer}$ \\
\hline
  A00   & 0  &   -  & -0.0453$\pm$0.0010 & -0.0387$\pm$0.0017 & -0.0554$\pm$0.0007   \\
\hline
  E01   & 6  &    0  & -0.0449$\pm$0.0007 & -0.0416$\pm$0.0008 & -0.0544$\pm$0.0010   \\
  E02   & 6  &   50  & -0.0404$\pm$0.0008 & -0.0364$\pm$0.0007 & -0.0531$\pm$0.0006   \\
  E03   & 6  &  100  & -0.0326$\pm$0.0006 & -0.0265$\pm$0.0007 & -0.0426$\pm$0.0007   \\
  E04   & 6  &  150  & -0.0300$\pm$0.0008 & -0.0231$\pm$0.0008 & -0.0460$\pm$0.0014   \\
  E05   & 6  &  200  & -0.0298$\pm$0.0007 & -0.0223$\pm$0.0004 & -0.0439$\pm$0.0018   \\
\tableline\end{tabular}
\tablecomments{First column: Model series number. 
Column 2: Satellite mass in units of solar masses.
Column 3: Initial velocity of the satellite in km~s$^{-1}$. 
Column 4: Oxygen abundance gradients from fitting radial abundance
profiles across the disk radius range of $R = 0$ to 25 kpc in dex~kpc$^{-1}$.
Column 5: Oxygen abundance gradients from fitting radial abundance
profiles in the inner disk ($R = 5$ -- 15 kpc) in dex~kpc$^{-1}$.
Column 6: Oxygen abundance gradients from fitting radial abundance
profiles in the outer disk ($R = 15$ -- 25 kpc) in dex~kpc$^{-1}$.
The first row of this table shows the results for a host galaxy without a satellite.}
\end{center}
\end{table*}

\begin{figure}[!htb]
  \resizebox{\hsize}{!}{\includegraphics{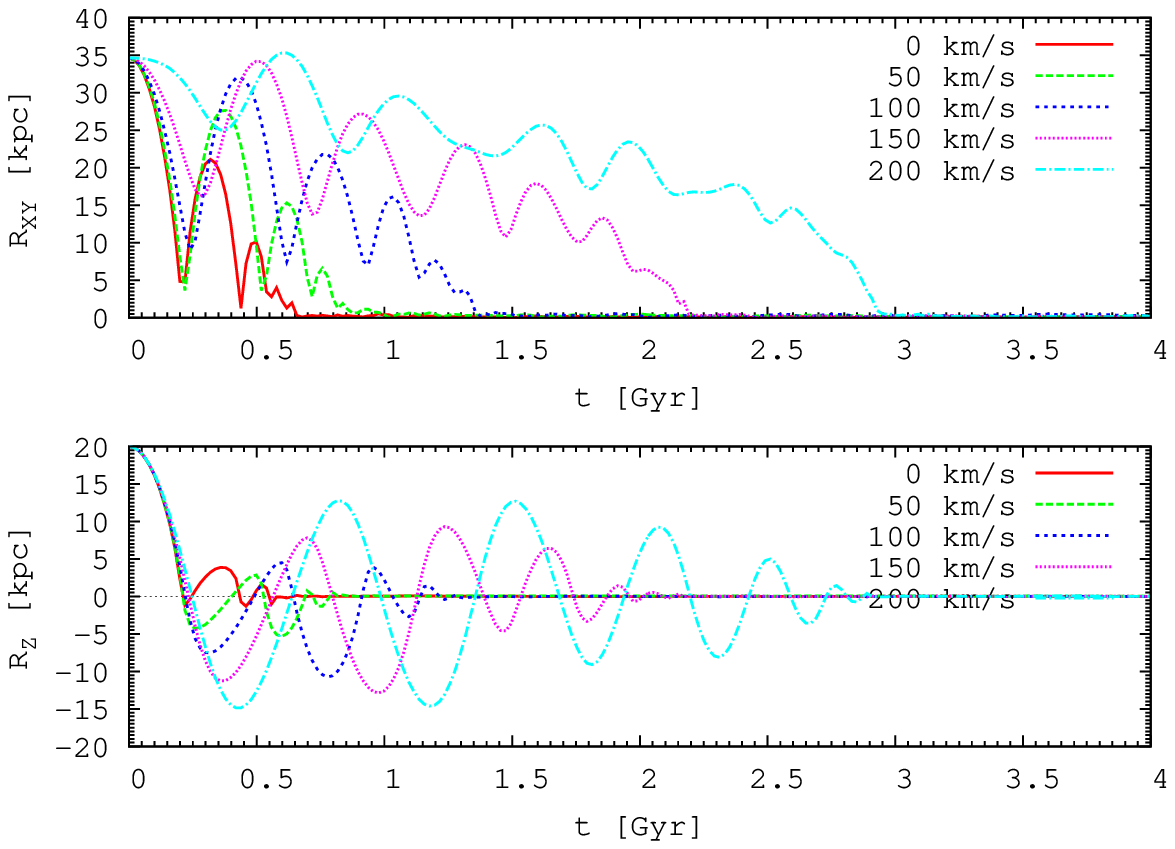}}
  \caption{Orbital evolution of satellites in models that have the same 
	satellite initial positions $R_{\rm XY} = 35$~kpc and $R_{\rm Z} = 20$~kpc (initial orbital 
  inclination $i = 30^\circ$) and mass ($6 \times 10^{9}$ M$_{\odot}$) 
	but different initial velocity. Top panel: evolution of distance from the 
	host galaxy center to the satellite in the disk plane. Bottom panel: evolution 
	of satellite height above the host galaxy disk plane.}
  \label{fig_dist-e}
\end{figure}

Figure~\ref{fig_dist-e} shows the orbital evolution of satellites for
this set of models. The top panel shows the evolution of distance from
the host galaxy to the satellite in the disk plane. The bottom panel
shows the evolution of satellite height above the host galaxy disk
plane. The times needed for settling the satellite in the host galaxy
center vary from $\sim 0.7$~Gyr (initial satellite velocity equals $V
= 0$~km~s$^{-1}$) to $\sim 3$~Gyr (satellite velocity $V =
200$~km~s$^{-1}$).

\begin{figure}[!htb]
  \resizebox{\hsize}{!}{\includegraphics{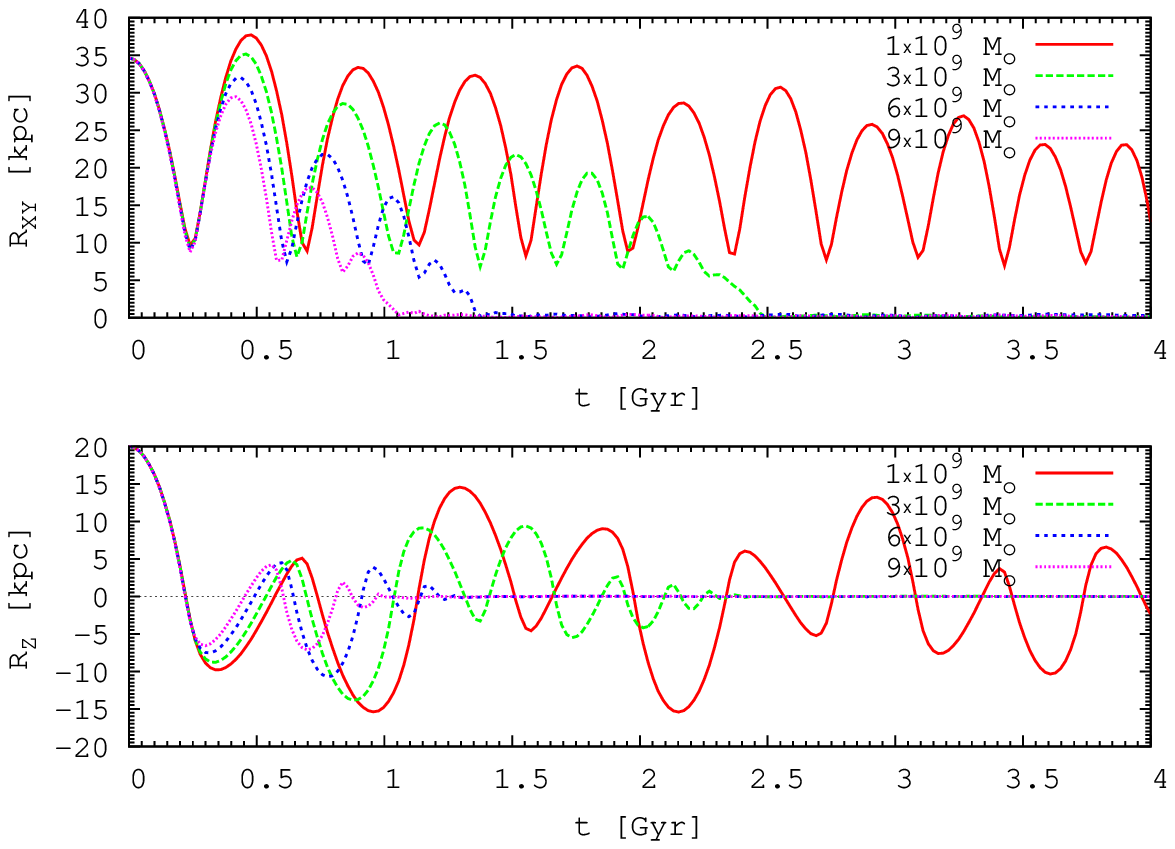}}
  \caption{Orbital evolution of satellites in our models A02, B02, C02, D02, which 
	have the same satellite initial positions  $R_{\rm XY} = 35$~kpc and $R_{\rm Z} = 20$~kpc (initial 
	orbital inclination $i = 30^\circ$)	and masses from $10^{9}$ M$_{\odot}$ 
	to $9 \times 10^{9}$ M$_{\odot}$. Top panel: evolution of the distance from 
	the host galaxy center to the satellite in the disk plane. Bottom panel: evolution of 
	satellite height above the host galaxy disk plane.}
  \label{fig_dist}
\end{figure}

Figure~\ref{fig_dist} shows the orbital evolution of satellites in the
models A02, B02, C02, and D02, which have the same satellite initial
position $R_{\rm XY} = 35$~kpc and $R_{\rm Z} = 20$~kpc (initial
orbital inclination $i = 30^\circ$) and masses from $10^{9}$
M$_{\odot}$ to $9 \times 10^{9}$ M$_{\odot}$. The top panel shows the
evolution of distance from the host galaxy to the satellite in the
disk plane. The bottom panel shows the evolution of satellite height
above the host galaxy disk plane.

The two right columns of Figure~\ref{fig_image} show
disk+bulge+satellite column density for the model with a satellite
mass of $6 \times 10^{9}$ M$_{\odot}$ (mass ratio 1:12), initial
satellite position $R_{\rm XY} = 35$~kpc and $R_{\rm Z} = 20$~kpc
(initial orbital inclination $i = 30^\circ$) and velocity $V =
100$~km~s$^{-1}$ at the times $t = 0$, 0.4, 0.8, 1.2, and 4~Gyr. For
comparison, the two left columns of Figure~\ref{fig_image} show the
column density for the run without satellite. The bottom right panel
of Figure~\ref{fig_oh-grad-evol} illustrates the evolution of the
radial metallicity distribution in the host galaxy during the merger
with the satellite.  The solid black line shows the initial
metallicity distribution.  The dotted lines 
mark the metallicity distributions at different times up to
4~Gyr. The top right panel of the Figure~\ref{fig_oh-grad-evol} shows
the evolution of the metallicity distribution in the galaxy without
satellite. 

The evolution of the metallicity distribution in the model C02 can be
divided into two stages.  During the first stage, for up to $t \approx
1.2$~Gyr, the satellite moves along a semi-periodic orbit around the
host galaxy, perturbing the disk of the host galaxy. These
perturbations induce the radial migration of stars and, therefore,
result in changes in the metallicity distribution across the disk,
increasing the abundances in its outer region ($R \gtrsim 10$~kpc).
During the second stage,  the satellite approaches the host galaxy and
passes through  denser regions. The rates of loss of both angular
momentum and kinetic energy increase due to the intense dynamical
friction \citep[see, e.g.,][]{JKBES2011}. This effect moves the
satellite towards the disk and then into the center of the host
galaxy: in model C02, the surface density defined by the disk
particles at the end of the simulation ($t = 4$~Gyr) is more than 100
times larger than the surface density defined by the satellite
particles for all disk regions except the innermost region ($R <
5$~kpc). The lines on the bottom right panel of
Figure~\ref{fig_oh-grad-evol} show the radial abundance distributions
in the host galaxy at the times $t = 1.2$, 3, and 4~Gyr. There are no
significant changes in the oxygen abundances in the inner part of the
disk ($R < 6$~kpc), but the abundances at large galactocentric
distances increase by $\sim 0.2$~dex. The slope of the abundance
gradient at galactocentric distances $5 < R < 15$~kpc becomes flatter. 

Figure~\ref{fig_Lz-e} shows the angular momentum distribution
evolution for models that have the same satellite initial position
$R_{\rm XY} = 35$~kpc and $R_{\rm Z} = 20$~kpc (initial orbital
inclination $i = 30^\circ$) and mass $6 \times 10^{9}$~M$_{\odot}$),
but different values of the initial velocity. Both axes are in kpc
units due to the normalization of the angular momentum by the circular
velocity in the disks. All rows except the last represent the
successive changes in angular momentum within the time interval of
$\Delta t = 0.5$~Gyr.  The last row represents the total cumulative
changes of the angular momentum from 0 to 4~Gyr.

\begin{figure}[htb]
  \resizebox{\hsize}{!}{\includegraphics{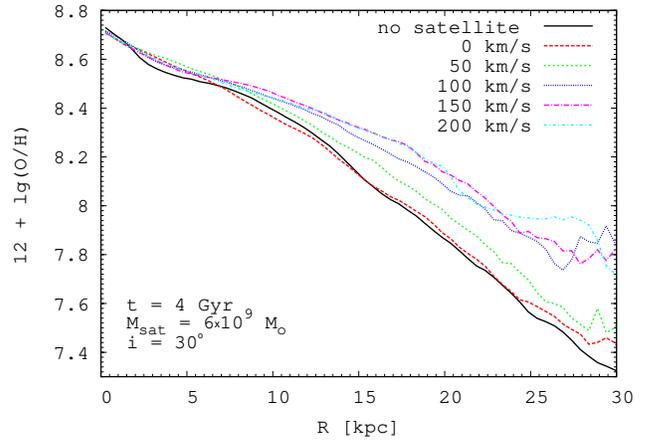}} 
  \caption{Radial oxygen abundance profile at 4~Gyr. The solid black line is a 
	model without satellite. The dotted lines indicate satellites with different initial 
	orbital velocities (models E01 -- E05).}
  \label{fig_oh-grad-evol-e}
\end{figure}

\begin{figure*}[htp]
  \vskip -1.0cm
  \begin{center}
  \includegraphics[width=1.00\linewidth]{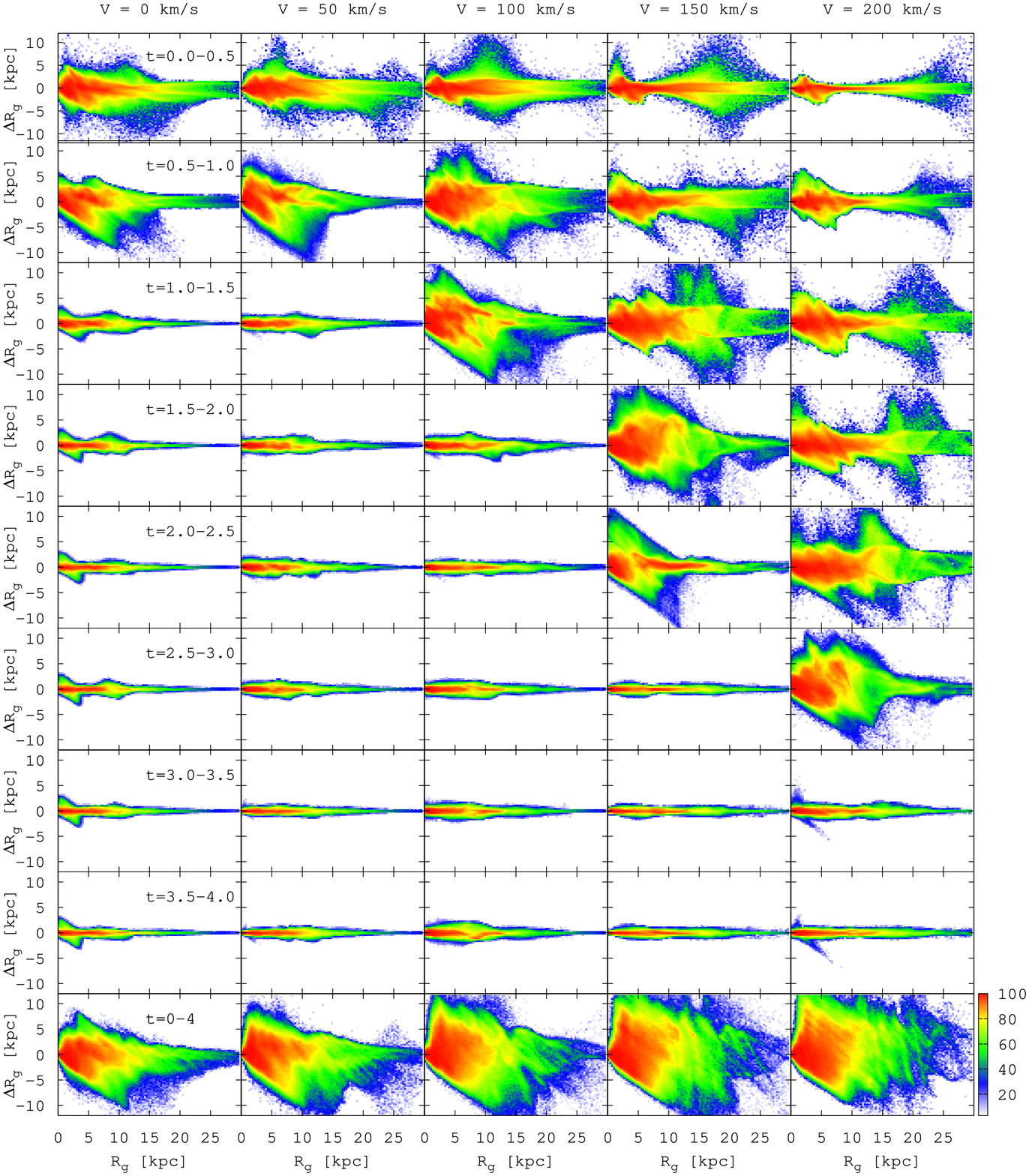} 
  \end{center}
  \vskip -0.5cm
  \caption{\footnotesize \baselineskip 0pt Angular momentum distribution 
	evolution for models that have the same satellite initial position 
	$R_{\rm XY} = 35$~kpc and $R_{\rm Z} = 20$~kpc (initial orbital inclination $i = 30^\circ$) 
	and mass $6 \times 10^{9}$ M$_{\odot}$ but different initial orbital 
	velocities (different columns). Both axes are in kpc units (see the text).
	Top to bottom: radial mixing in time intervals of $\Delta t =
0.5$~Gyr at 
	  (0 -- 0.5)~Gyr, (0.5 -- 1.0)~Gyr, (1.0 -- 1.5)~Gyr, (1.5
-- 2.0)~Gyr, 
	(2.0 -- 2.5)~Gyr, (2.5 -- 3.0)~Gyr, (3.0 -- 3.5)~Gyr, and
(3.5 -- 4.0)~Gyr.
        Last row: total between (0 -- 4)~Gyr.
        The color coding indicates the percentage of particles below some fixed levels (see color bar on the bottom right).}
  \label{fig_Lz-e}
\end{figure*}

Figure~\ref{fig_oh-grad-evol-e} shows the radial abundance profile at
$t = 4$~Gyr. The solid black line is the model without satellite. The
dotted lines represent models with
satellites of different initial velocities.  One can see in
Figure~\ref{fig_dist-e} and Figure~\ref{fig_Lz-e} that all the models
undergo  merging by $t = 4$~Gyr, so we can compare the influence of the
satellite's initial velocity on the metallicity gradient as the result
of the merging process.  The largest influence on the metallicity
gradient of the outer disk can be observed for the satellite initial
orbital velocity of $V \geq 100$~km~s$^{-1}$. A further increase of
the satellite's initial orbital velocity does not significantly change
the behavior of the radial oxygen abundance profile of the outer
part of the disk (Figure~\ref{fig_oh-grad-evol-e}).
Thus, we choose the satellite's initial orbital velocity $V =
100$~km~s$^{-1}$ as the basic setup for further investigation. 

\section{Results \& Discussion}

\subsection{Radial metallicity distribution}

Now we discuss the influence of the particle migration caused by the
merger of satellite and host galaxy with different mass ratios
and orbit orientations on the abundance distribution across the disk. 

\begin{figure*}[!ht]
   \resizebox{\hsize}{!}{\includegraphics{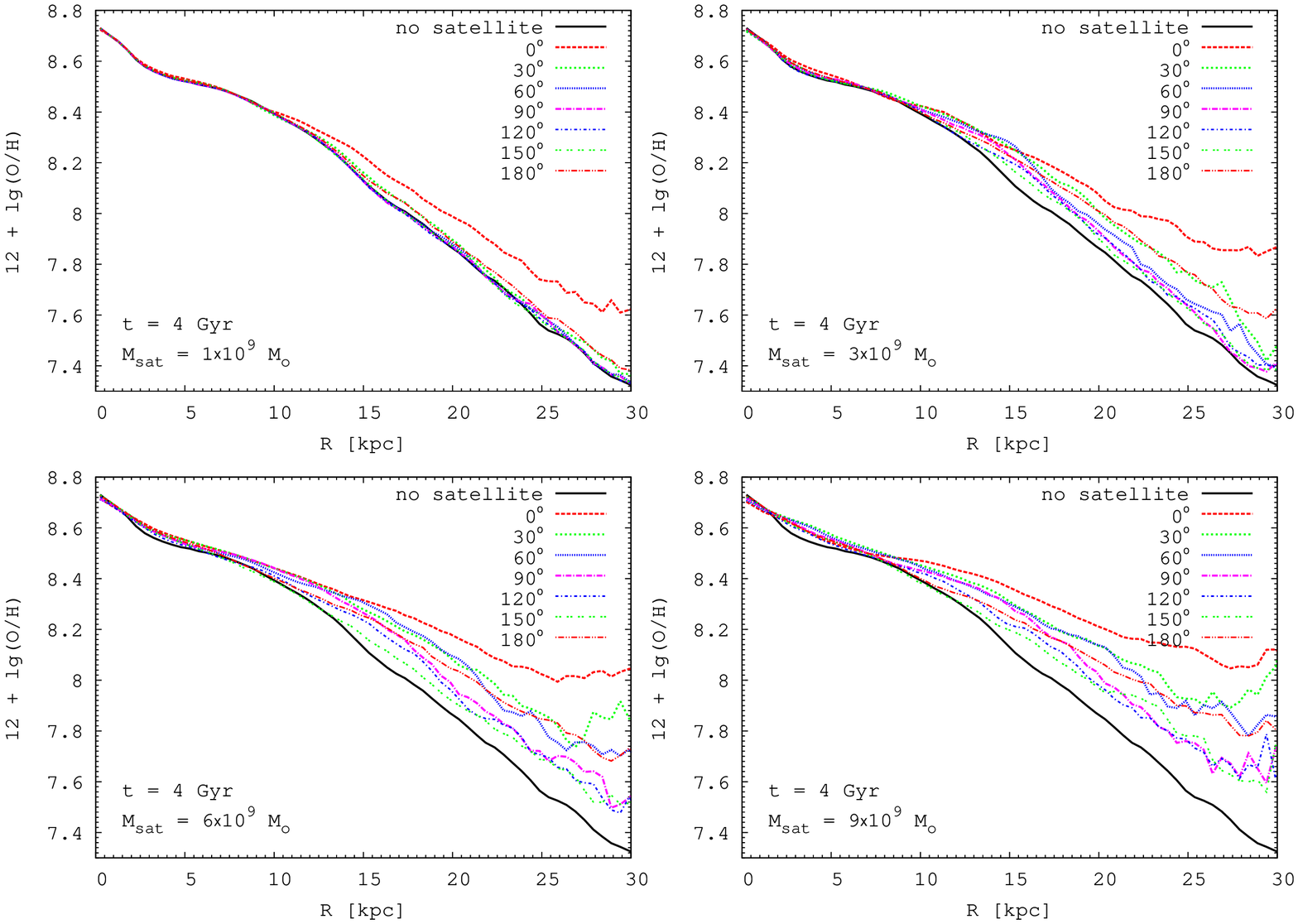}}
  \caption{Radial oxygen abundance profiles at 4~Gyr for a disk galaxy without 
	satellite (solid black line) and with satellite (dotted lines) after merging. 
  The satellite masses are $10^{9}$ M$_{\odot}$ (mass ratio 1:70) (top left panel),  
  $3 \times 10^{9}$ M$_{\odot}$ (mass ratio 1:23) (top right panel), 
  $6 \times 10^{9}$ M$_{\odot}$ (mass ratio 1:12) (bottom left panel), and 
  $9 \times 10^{9}$ M$_{\odot}$ (mass ratio 1:8) (bottom right panel).}
  \label{fig_oh-grad-sat1}
\end{figure*}

Figure~\ref{fig_oh-grad-sat1} shows the radial abundance distributions
at $t = 4$~Gyr.  The masses of the satellites are $10^{9}$ M$_{\odot}$
(upper left panel), $3 \times 10^{9}$ M$_{\odot}$ (upper right panel),
$6 \times 10^{9}$ M$_{\odot}$ (bottom left panel), and $9 \times
10^{9}$ M$_{\odot}$ (bottom right panel). The solid black line
corresponds to the evolution of the model without satellite, the
dotted lines show the host galaxy
accreting satellites with different initial orbit inclinations.  The
comparison between the model without and with satellites shows that
there are no significant changes ($\gtrsim 0.1$~dex) in the radial
abundance distribution after 4~Gyr of evolution in the inner part of
the disk of the host galaxy for all the satellite masses considered
(Figure~\ref{fig_oh-grad-sat1}).  In the case of the smallest
satellite mass ($10^{9}$~M$_{\odot}$) (top left panel), there are no
significant changes in the radial abundance distribution across the
whole disk except for the case when the orbital inclination of the
satellite $i = 0^\circ$. Accretion of more massive satellites causes a
significant increase (up to $\gtrsim 0.2 - 0.5$~dex) in the oxygen
abundances at galactocentric distances $R \gtrsim 10$~kpc, as
illustrated in the top right and bottom panels of
Figure~\ref{fig_oh-grad-sat1}. 

\begin{figure*}[!ht]
  \resizebox{1.02\hsize}{!}{\includegraphics{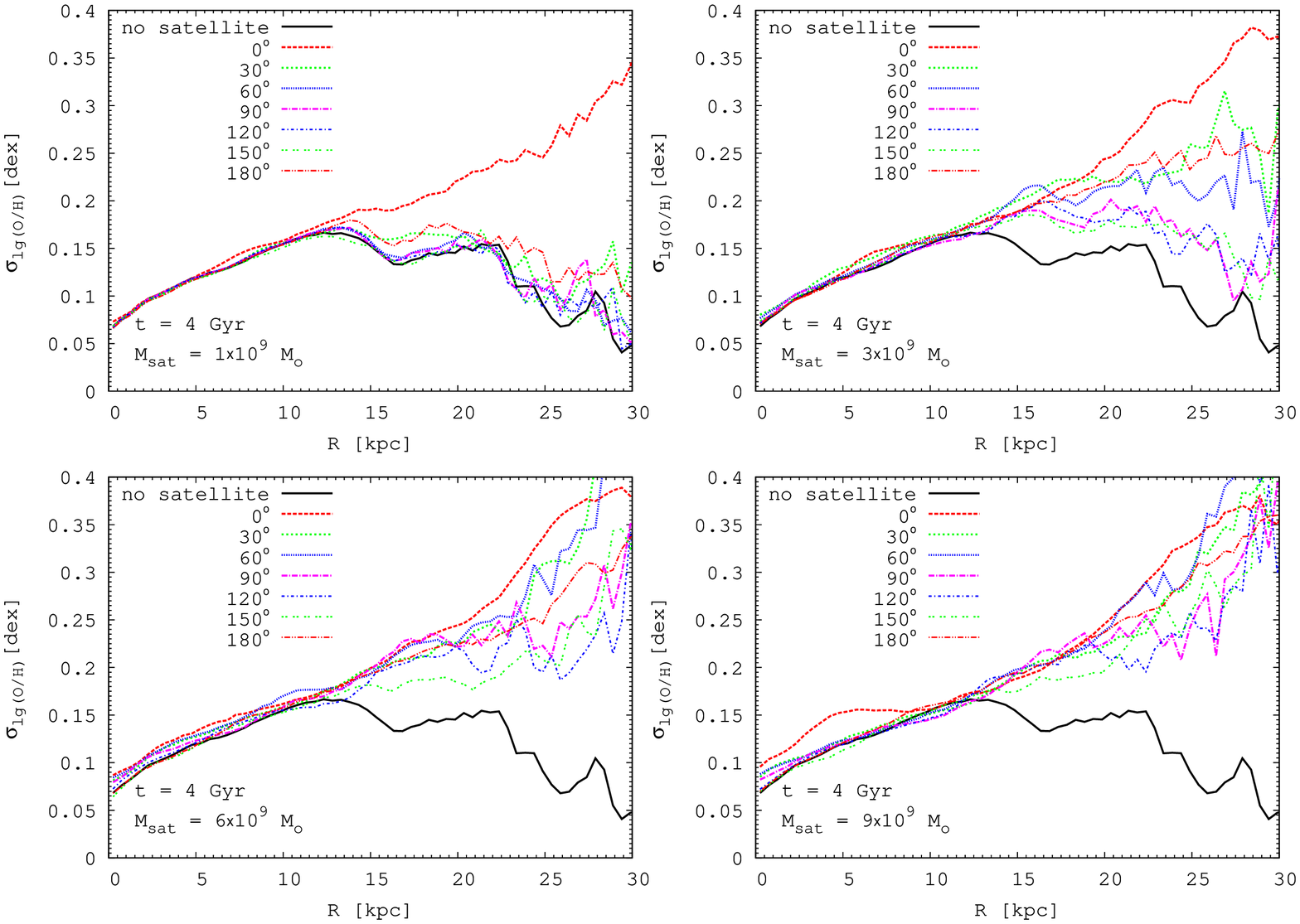}}
  \caption{Radial profiles of the oxygen abundance's standard deviation 
	$\sigma_{\lg({\rm O/H})}$ at 4~Gyr for a disk galaxy without satellite 
	(solid black line) and with satellite (dotted lines) after merging. 
  The satellite masses are $10^{9}$ M$_{\odot}$ (mass ratio 1:70) (top left panel),  
  $3 \times 10^{9}$ M$_{\odot}$ (mass ratio 1:23) (top right panel), 
  $6 \times 10^{9}$ M$_{\odot}$ (mass ratio 1:12) (bottom left panel), and 
  $9 \times 10^{9}$ M$_{\odot}$ (mass ratio 1:8) (bottom right panel).}
  \label{fig_sigma2}
\end{figure*}

The accretion of a satellite with a mass of $M_{sat} \gtrsim 3 \times
10^{9}$~M$_{\odot}$ or less with an orbit in the disk plane in
prograde direction can lead to a significant increase of the
metallicity in the outer regions of the host galaxy. It should be
noted that the  abundance redistribution due to merging strongly
depends not only on the satellite mass but also on its initial orbital
inclination. The maximum redistribution is caused by satellites with
prograde orbits ($i = 0^\circ$) in the disk plane of the galaxy. The
minimum redistribution is caused by satellites with retrograde orbits
with an initial inclination $i = 90^\circ$ to $150^\circ$.

Figure~\ref{fig_sigma2} shows the mean deviation around the radial
abundance trend $\sigma_{\lg({\rm O/H})}$ (in radial cylindrical bins)
at $t = 4$~Gyr for models without (solid black line) and with satellite
(dotted lines). Models with the four
satellite masses 
$10^{9}$~M$_{\odot}$ (mass ratio 1:70) (top left panel), 
$3 \times 10^{9}$ M$_{\odot}$ (mass ratio 1:23) (top right panel), 
$6 \times 10^{9}$ M$_{\odot}$ (mass ratio 1:12) (bottom left panel),
and
$9 \times 10^{9}$ M$_{\odot}$ (mass ratio 1:8) (bottom right panel) 
are presented.  The scatter in the abundances at a given radius caused
by radial migration of the particles in the model without satellite
increases with galactocentric radius up to $R \sim 15$~kpc and
decreases beyond this radius.  The scatter in metallicities at a given
radius in the outer disk ($\gtrsim$ 15~kpc) significantly increases
for most satellite initial masses/positions compared to the isolated
galaxy case.

\begin{figure}[ht]
  \resizebox{1.08\hsize}{!}{\includegraphics{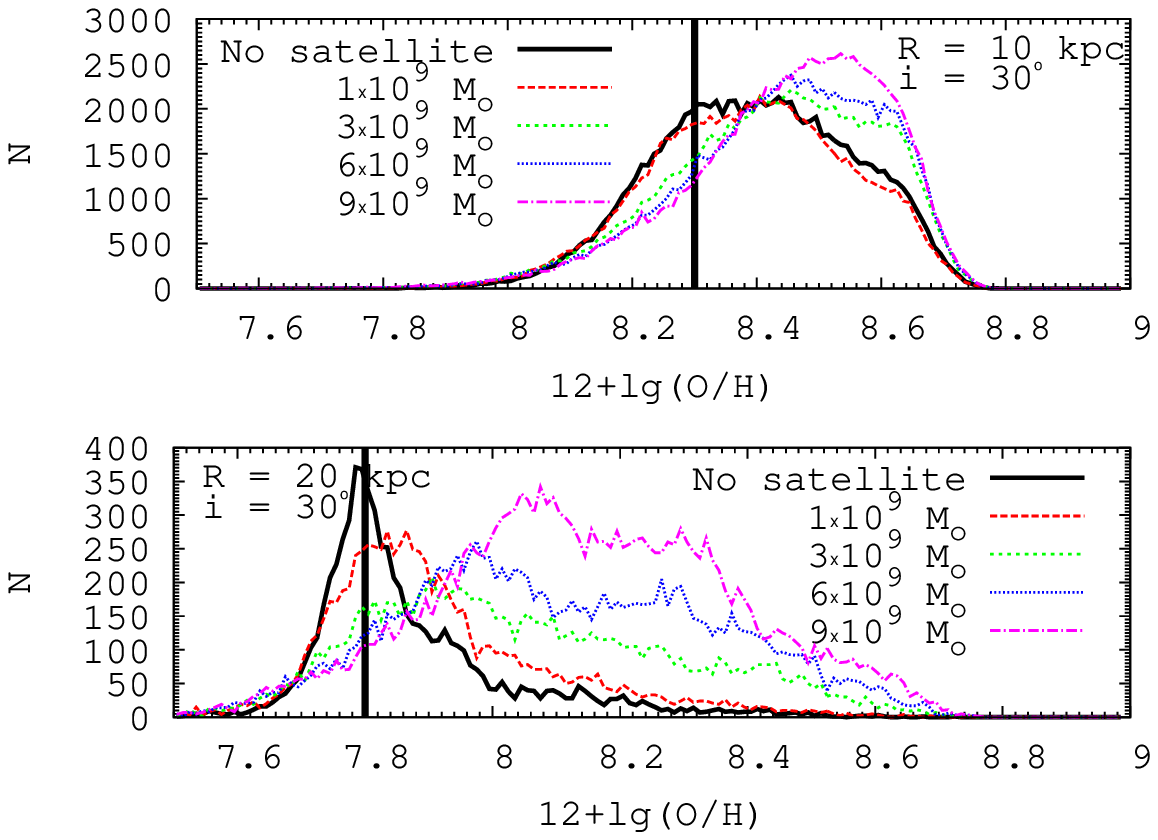}}
  \caption{Oxygen abundance distribution histogram at 4~Gyr for galactocentric 
	radii of 10~kpc (top panel) and 20~kpc (bottom panel) for models that 
	have the same satellite initial orbital inclination $i$ =
30$^\circ$ and velocity $V = 100$~km~s$^{-1}$
	but different masses. The vertical black line indicates the initial abundance at 
	the galactocentric radii of 10~kpc (top panel) and 20~kpc (bottom panel).} 
  \label{fig_oh-slice-M}
\end{figure}

\begin{figure}[ht]
  \resizebox{1.08\hsize}{!}{\includegraphics{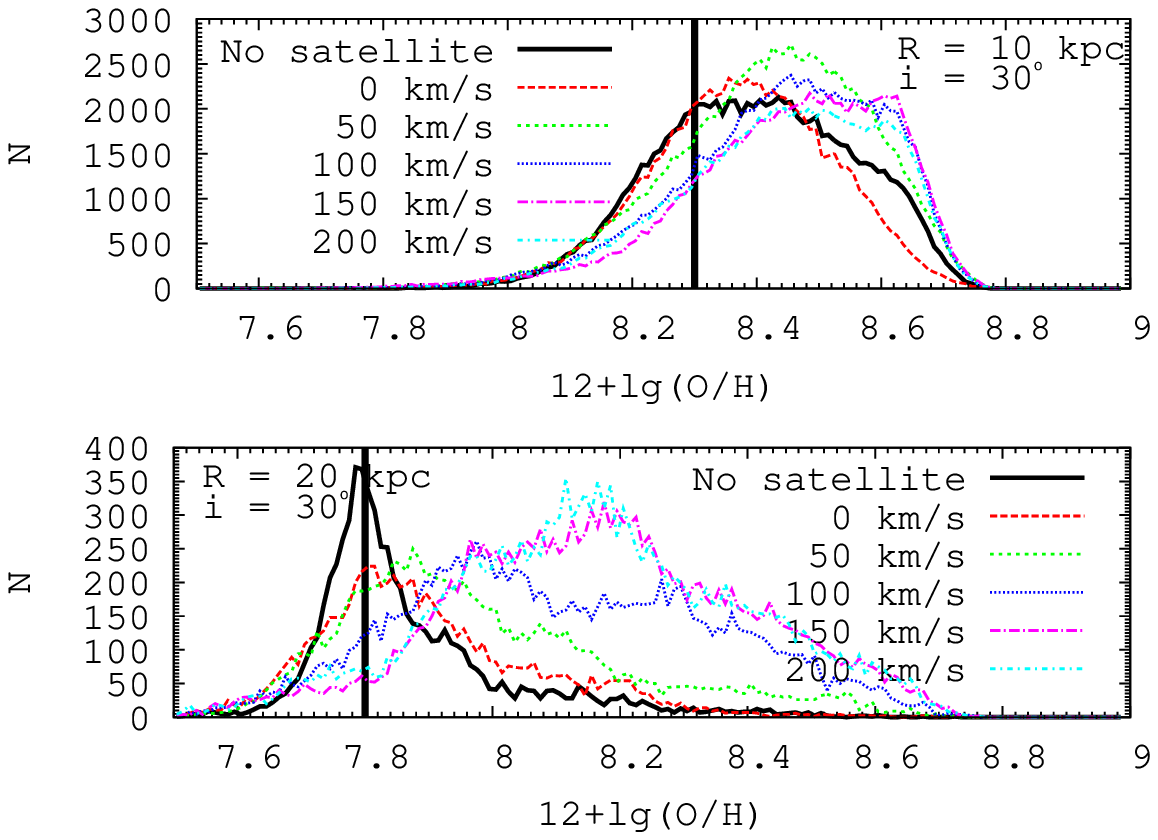}}
  \caption{Oxygen abundance distribution histogram at 4~Gyr for galactocentric 
	radii of 10~kpc (top panel) and 20~kpc (bottom panel) for models that 
	have the same satellite initial orbital inclination $i$ =
30$^\circ$ and a mass of $6 \times 10^{9}$ M$_{\odot}$ 
	but different initial velocity. The vertical black line marks initial abundance at 
	the galactocentric radii of 10~kpc (top panel) and 20~kpc (bottom panel).} 
  \label{fig_oh-slice-e}
\end{figure}

Figure~\ref{fig_oh-slice-M} shows the abundance distribution histogram
for particles at $t = 4$~Gyr for the models A00 (without satellite)
and A02, B02, C02, D02 (with satellites of different masses) at
galactocentric radii of $R = 10$~kpc (top panel) and $R = 20$~kpc
(bottom panel). The vertical black lines show the initial abundance
distributions at these radii. It is clear that the accretion of a
satellite with a mass $\gtrsim 3 \times 10^{9}$~M$_{\odot}$
significantly shifts the abundance distribution at both galactocentric
distances to higher values and the shift is larger at larger radii. A
small fraction of particles with high abundances from the central
region migrates to the outer disk region even in the case of an
isolated galaxy. These particles produce an asymmetric particle
abundance distribution with a high abundance tail (see the solid black
line at the bottom panel in Figure~\ref{fig_oh-slice-M}).
Figure~\ref{fig_oh-slice-e} shows again the abundance distribution
histogram for particles at 4~Gyr for galactocentric radii of 10~kpc
(top panel) and 20~kpc (bottom panel) predicted by the models with a
fixed initial satellite orbit inclination ($i = 30^\circ$) and mass
(M$_{sat}$ = 6 $\times$ 10$^{9}$ M$_{\odot}$) but with different
initial velocities from $V = 0$ to 200~km~s$^{-1}$.

\subsection{Radial metallicity distribution breaks}

Our simulations provide radial metallicity distributions in the outer
disk (up to 30~kpc) for a Milky Way-like disk galaxy. It gives us an
opportunity to test whether the shallow abundance gradient observed in
the outer disks of some spiral galaxies can be attributed to star
particle migration caused by merging. The part of the disk with $R$
between 5 and 15~kpc will be referred to as the inner (optical) disk
while the part of the disk between 15 and 25~kpc will be referred to
as the outer (extended) disk.  The innermost region $R \lesssim 5$~kpc
is not discussed here.  The region beyond $R = 25$~kpc is also
excluded from our discussion since the number of particles there is
not high enough for reliable estimations of metallicities. 

A close examination of the radial metallicity distributions shows that
the abundance distribution slopes slightly change with radius. To
quantify the variations in the abundance gradient slopes we fit the
abundance distributions at 4~Gyr with separate exponentials in the
inner and outer disks for each simulation from Table~\ref{table1} and
Table~\ref{table2}.  The obtained slopes are listed in
Table~\ref{table2}. The first column gives the model number.  Column 2
reports the satellite mass, column 3 lists the initial distance of the
satellite, column 4 reports the initial satellite inclination, columns
5, 6, and 7 give the slope of abundance gradient in entire disk ($R =
0$ to 25~kpc), in the inner disk ($R = 5$ to 15~kpc), and in the outer
disk ($R = 15$ to 25~kpc), respectively.

Inspection of Table~\ref{table2} shows that the slopes of the
abundance gradient at $t = 4$~Gyr in the inner disk
($-0.039$~dex~kpc$^{-1}$) and in the outer disk
($-0.055$~dex~kpc$^{-1}$) of an isolated galaxy (model A00) are quite
close to the initial slope of $-0.05$~dex~kpc$^{-1}$.  The slopes of the
gradient in the outer (extended) disk in the majority of the models
with satellites are  close to that in the model of an isolated galaxy.
In contrast, the slopes of the gradient  in the inner disk in the
majority of the models with satellites of masses $\gtrsim 3 \times
10^9$~M$_{\odot}$ decrease. 
Assuming a mass-to-light ratio in the B band of $M/L_B = 1.4$
\citep{Flynn2006}, the optical ($R_{25}$) radius of our host galaxy is
$\sim 15$~kpc.  Thus, the flattening of the gradient observed in
extended ultraviolet disks (XUV-disk) of some spiral galaxies
\citep{GildePaz2007,Thilker2007} cannot be attributed to particle
migration caused by the accretion of a low-mass satellite at least for
Milky Way-like type galaxies.

\begin{table*}[ht]
\footnotesize 
\begin{center}
\caption{Model runs with different satellite masses and initial positions and the resulting disk oxygen abundance gradients.}
\label{table2}
\begin{tabular}{c|ccrcccccc}
\tableline
  Model & Mass [$\times10^{9} M_{\odot}$] & $R$ [kpc] & $i$ [deg] & $\Delta$$\lg$(O/H)$_{\rm all}$ & $\Delta$$\lg$(O/H)$_{\rm inner}$ & $\Delta$$\lg$(O/H)$_{\rm outer}$ \\
\hline  
  A00   & 0  &  - &   - & -0.0453$\pm$0.0010 & -0.0387$\pm$0.0017 & -0.0554$\pm$0.0007  \\ 
\hline
  A01   & 1  & 40 &   0 & -0.0375$\pm$0.0007 & -0.0303$\pm$0.0006 & -0.0466$\pm$0.0006  \\ 
  A02   & 1  & 40 &  30 & -0.0439$\pm$0.0010 & -0.0361$\pm$0.0012 & -0.0573$\pm$0.0007  \\ 
  A03   & 1  & 40 &  60 & -0.0449$\pm$0.0009 & -0.0387$\pm$0.0017 & -0.0535$\pm$0.0008  \\ 
  A04   & 1  & 40 &  90 & -0.0451$\pm$0.0009 & -0.0385$\pm$0.0018 & -0.0527$\pm$0.0007  \\ 
  A05   & 1  & 40 & 120 & -0.0454$\pm$0.0010 & -0.0386$\pm$0.0018 & -0.0538$\pm$0.0005  \\ 
  A06   & 1  & 40 & 150 & -0.0455$\pm$0.0010 & -0.0389$\pm$0.0017 & -0.0558$\pm$0.0008  \\ 
  A07   & 1  & 40 & 180 & -0.0434$\pm$0.0009 & -0.0371$\pm$0.0015 & -0.0533$\pm$0.0005  \\ 
\hline
  B01   & 3  & 40 &   0 & -0.0315$\pm$0.0004 & -0.0274$\pm$0.0009 & -0.0353$\pm$0.0011  \\ 
  B02   & 3  & 40 &  30 & -0.0364$\pm$0.0009 & -0.0262$\pm$0.0008 & -0.0519$\pm$0.0008  \\ 
  B03   & 3  & 40 &  60 & -0.0400$\pm$0.0012 & -0.0257$\pm$0.0002 & -0.0603$\pm$0.0011  \\ 
  B04   & 3  & 40 &  90 & -0.0416$\pm$0.0011 & -0.0297$\pm$0.0006 & -0.0596$\pm$0.0005  \\ 
  B05   & 3  & 40 & 120 & -0.0419$\pm$0.0010 & -0.0349$\pm$0.0011 & -0.0592$\pm$0.0006  \\ 
  B06   & 3  & 40 & 150 & -0.0427$\pm$0.0010 & -0.0347$\pm$0.0012 & -0.0567$\pm$0.0007  \\ 
  B07   & 3  & 40 & 180 & -0.0365$\pm$0.0006 & -0.0314$\pm$0.0006 & -0.0466$\pm$0.0005  \\ 
\hline
  C01   & 6  & 40 &   0 & -0.0265$\pm$0.0002 & -0.0240$\pm$0.0003 & -0.0310$\pm$0.0005  \\ 
  C02   & 6  & 40 &  30 & -0.0326$\pm$0.0006 & -0.0265$\pm$0.0007 & -0.0426$\pm$0.0007  \\ 
  C03   & 6  & 40 &  60 & -0.0331$\pm$0.0009 & -0.0242$\pm$0.0004 & -0.0516$\pm$0.0016  \\ 
  C04   & 6  & 40 &  90 & -0.0385$\pm$0.0011 & -0.0270$\pm$0.0011 & -0.0576$\pm$0.0008  \\ 
  C05   & 6  & 40 & 120 & -0.0391$\pm$0.0009 & -0.0302$\pm$0.0007 & -0.0546$\pm$0.0013  \\ 
  C06   & 6  & 40 & 150 & -0.0411$\pm$0.0006 & -0.0363$\pm$0.0007 & -0.0490$\pm$0.0005  \\ 
  C07   & 6  & 40 & 180 & -0.0342$\pm$0.0004 & -0.0300$\pm$0.0002 & -0.0431$\pm$0.0005  \\ 
\hline
  D01   & 9  & 40 &   0 & -0.0228$\pm$0.0004 & -0.0154$\pm$0.0006 & -0.0247$\pm$0.0010  \\ 
  D02   & 9  & 40 &  30 & -0.0302$\pm$0.0005 & -0.0243$\pm$0.0004 & -0.0362$\pm$0.0011  \\ 
  D03   & 9  & 40 &  60 & -0.0316$\pm$0.0006 & -0.0256$\pm$0.0005 & -0.0429$\pm$0.0018  \\ 
  D04   & 9  & 40 &  90 & -0.0370$\pm$0.0011 & -0.0239$\pm$0.0005 & -0.0570$\pm$0.0009  \\ 
  D05   & 9  & 40 & 120 & -0.0380$\pm$0.0008 & -0.0302$\pm$0.0013 & -0.0505$\pm$0.0010  \\ 
  D06   & 9  & 40 & 150 & -0.0372$\pm$0.0004 & -0.0365$\pm$0.0004 & -0.0395$\pm$0.0015  \\ 
  D07   & 9  & 40 & 180 & -0.0325$\pm$0.0002 & -0.0303$\pm$0.0004 & -0.0361$\pm$0.0005  \\ 
\tableline\end{tabular}
\tablecomments{First column: Model series number. 
Column 2: Satellite mass in units of solar masses.
Column 3: Initial distance of the satellite from the center of the disk galaxy.
Column 4: Initial orbital inclination of the satellite.
Column 5: Oxygen abundance gradients from fitting radial abundance
profile across the disk radius range of $R = 0$ to 25 kpc in dex~kpc$^{-1}$.
Column 6: Oxygen abundance gradients from fitting radial abundance
profiles in the inner disk ($R = 5$ -- 15 kpc) in dex~kpc$^{-1}$.
Column 7: Oxygen abundance gradients from fitting radial abundance
profiles in the outer disk ($R = 15$ -- 25 kpc) in dex~kpc$^{-1}$.
The first row of this table shows the results for a host galaxy without a satellite.}
\end{center}
\end{table*}

\subsection{Particle migration}

\begin{figure*}[htb]
  \resizebox{0.95\hsize}{!}{\includegraphics{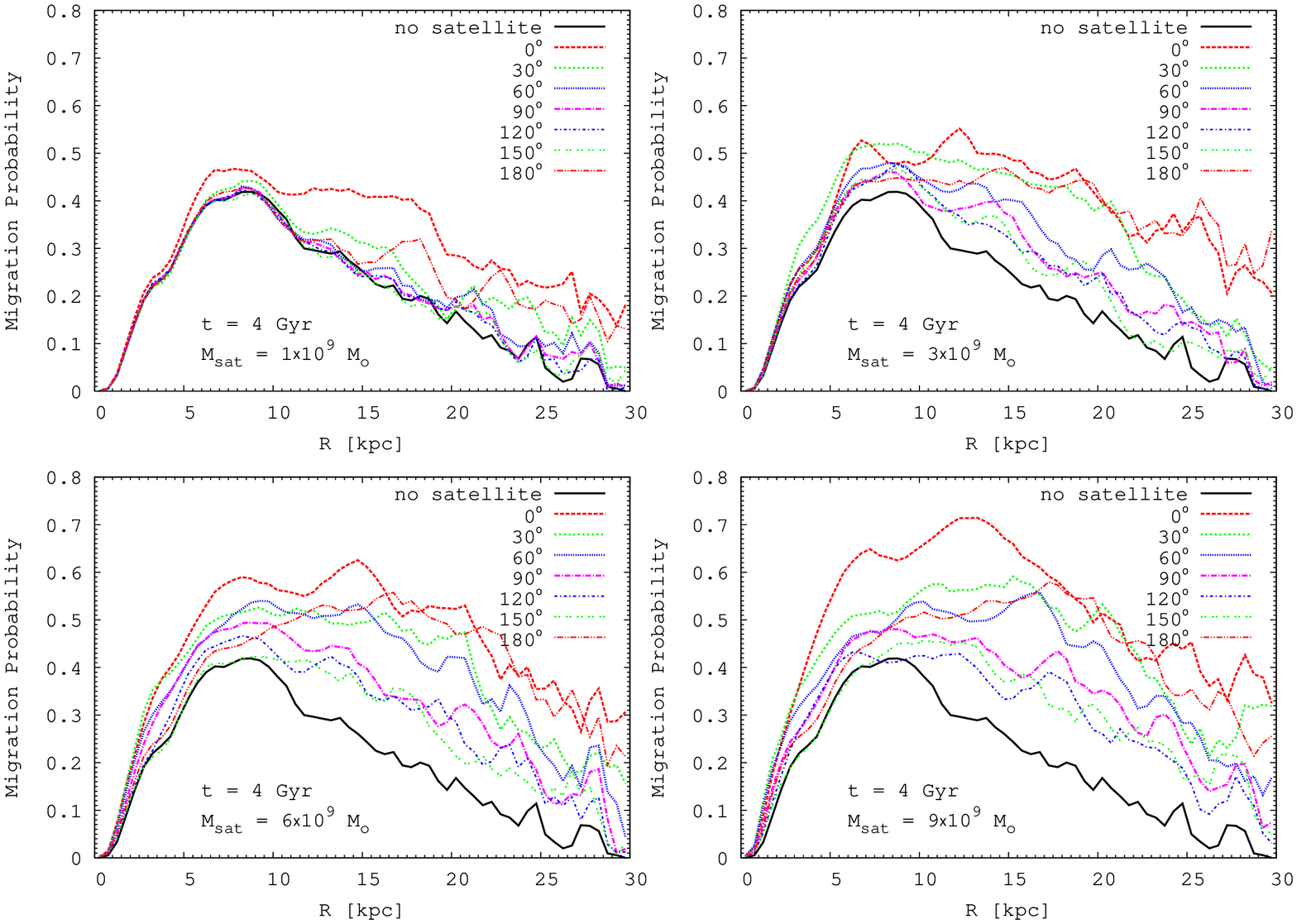}}
  \caption{The migration probability of disk and bulge particles. The lines correspond to the fraction of particles that 
  change their galactocentric radius by more than 3~kpc after 4~Gyr of evolution.}
  \label{fig_migrprob}
\end{figure*}

\citet{Bird2012} noticed that the processes of stellar migration
induced by accretion of a population of dark matter sub-halos are
distinct from those in an isolated galactic disk. They found that the
migration probability of particles traces the radial mass distribution
in the case of an isolated disk while in the perturbed disk this
correlation is absent.  Our simulations reveal that the migration
probability of particles in a perturbed disk depends on the mass and
orbital parameters of the merged satellite
(Figure~\ref{fig_migrprob}). The migration probability of disk
particles correlates with disk surface density beyond $R\sim 7$~kpc
for the case of an isolated and some of the disturbed disks (see
Figure~\ref{fig_migrprob} and top left panel of
Figure~\ref{fig_oh-grad-evol} for a comparison with surface density).
Such radial profiles of the particle migration probability are similar
to the one assumed in the chemical evolution model of our Galaxy of
\citet{Schonrich2009}. For the case of some other mass and orbital
parameters of the merged satellite (e.g., when the satellite's orbit
lies in the disk plane) the migration probability of particles remains
about constant at $R\sim 7$ to 20~kpc as it found by \citet{Bird2012},
but decreases with radius beyond $R\sim 20$~kpc.

\begin{figure*}[htbp]
  \vskip -1.0cm
  \begin{center}
  \includegraphics[width=1.00\linewidth]{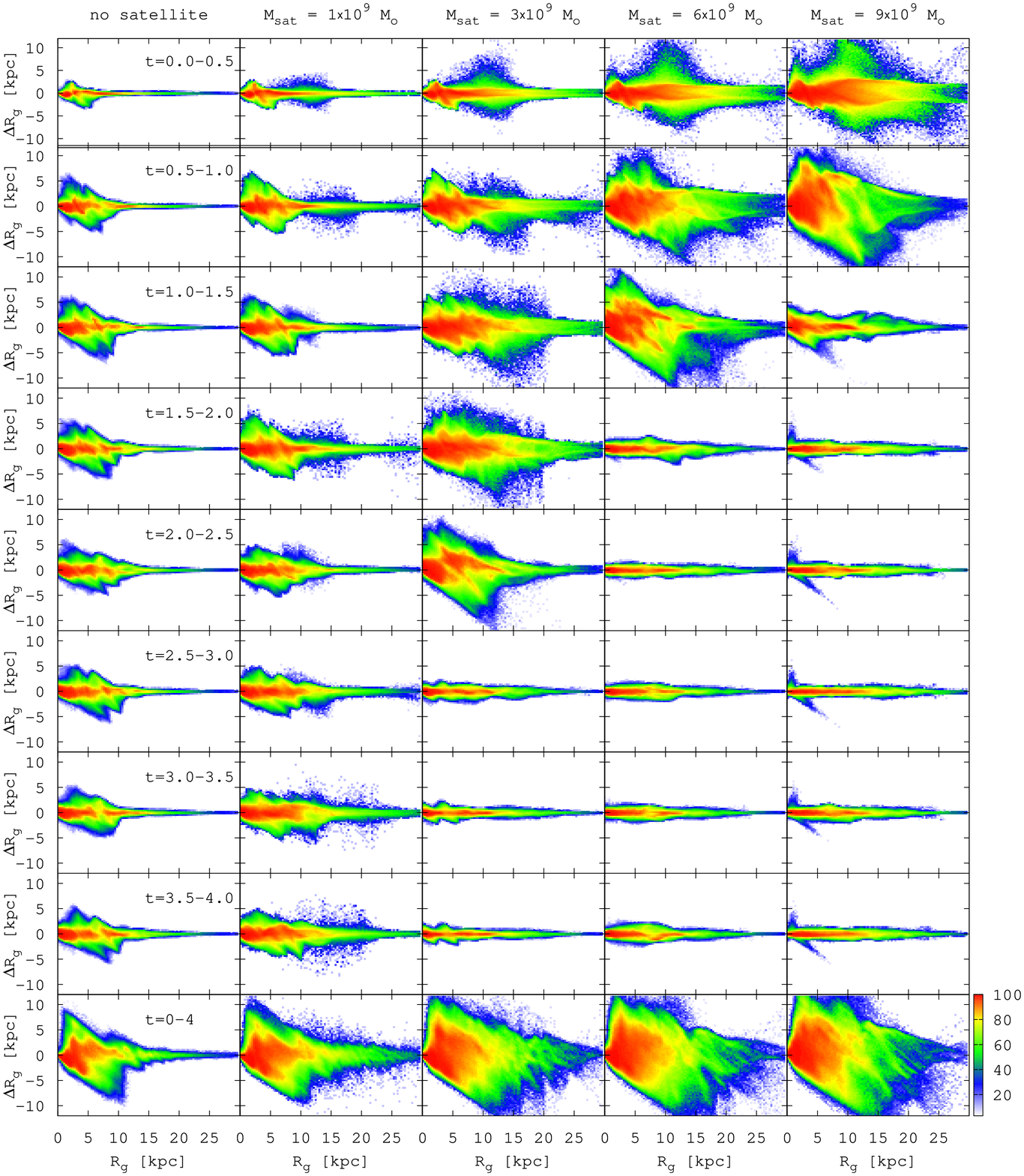} 
  \end{center}
  \vskip -0.5cm
  \caption{\footnotesize \baselineskip 0pt Comparison of the angular momentum 
	distribution for the model A00 (without satellite) and the models A02, B02, C02, D02 
	(with different satellite masses) in the columns from the left to the right. Both axes are in kpc units 
	(see the text). Top to bottom: radial mixing in time intervals
of $\Delta t = 0.5$~Gyr 
	at (0 -- 0.5)~Gyr, (0.5 -- 1.0)~Gyr, (1.0 -- 1.5)~Gyr, (1.5 -- 2.0)~Gyr, 
	(2.0 -- 2.5)~Gyr, (2.5 -- 3.0)~Gyr, (3.0 -- 3.5)~Gyr, and
(3.5 -- 4.0)~Gyr. Last row: 
	total between (0 -- 4)~Gyr.
	The color coding indicates the percentage of particles below some fixed levels (see color bar on the bottom right).}
  \label{fig_Lz-M}
\end{figure*}

\begin{figure*}[htbp]
  \vskip -1.0cm
  \begin{center}
  \includegraphics[width=1.00\linewidth]{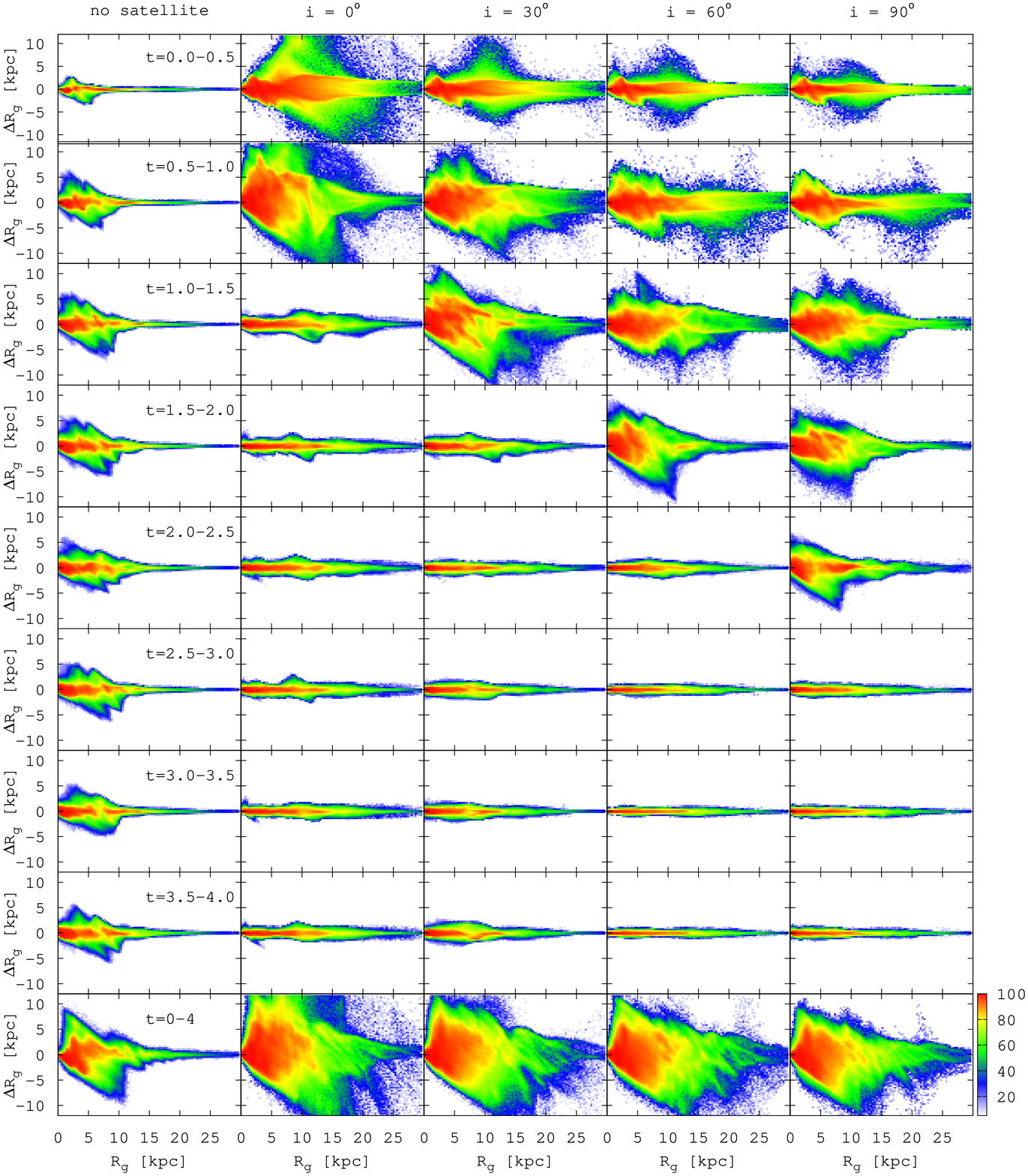} 
  \end{center}
  \vskip -0.5cm
  \caption{\footnotesize \baselineskip 0pt Comparison of the angular momentum 
	distribution for the model A00 (without satellite) and the
models C01 -- C04 
	(with different satellite initial orbital inclinations) in the columns from the left to the right. 
  Both axes are in kpc units (see the text). Top to bottom: radial mixing in 
	time intervals of $\Delta t = 0.5$~Gyr at (0 -- 0.5)~Gyr,
(0.5 -- 1.0)~Gyr, 
	(1.0 -- 1.5)~Gyr, (1.5 -- 2.0)~Gyr, (2.0 -- 2.5)~Gyr, (2.5
-- 3.0)~Gyr, 
	(3.0 -- 3.5)~Gyr, and (3.5 -- 4.0)~Gyr. Last row: total
between (0 -- 4)~Gyr.
	The color coding indicates the percentage of particles below some fixed levels (see color bar on the bottom right).}
  \label{fig_Lz-i}
\end{figure*}

Figure~\ref{fig_Lz-M} shows the angular momentum distribution
evolution for models without satellite (A00) and with satellites of
different  masses from $10^{9}$ M$_{\odot}$ to $9 \times 10^{9}$
M$_{\odot}$ (A02, B02, C02, D02). Both axes are in kpc units due to
the normalization of the angular momentum by the circular velocity in
the disks. All rows except for the last one represent the successive
changes in angular momentum within the time interval of $\Delta t =
0.5$~Gyr. The last row represents the total cumulative changes of the
angular momentum from 0 to 4~Gyr.  Figure~\ref{fig_Lz-i} shows the
same as the previous figure but for the angular momentum distribution
evolution for models with a fixed satellite mass of $6 \times 10^{9}$
M$_{\odot}$ and for different initial orbital inclinations $i =
0^\circ$, $30^\circ$, $60^\circ$, and $90^\circ$ (models C01 -- C04).
When considering the angular momentum changes of disk particles after
merging (bottom rows of Figure~\ref{fig_Lz-M} and
Figure~\ref{fig_Lz-i}) we find that the maximal changes in guiding
radius $\Delta R_g$ decrease outwards even for models that show a
fairly constant value of migration probability within $R \sim 7$ to
20~kpc.  Those migration patterns lead to the number of particles
migrating to the outer disk region being higher than the number of
those migrating in the opposite direction. Regarding metallicity
profile changes this means an increase of the mean metallicity in the
outer disk region as is evident from our metallicity profile plots
discussed above.  The comparison of angular momentum changes in short
time bins of $\Delta t = 0.5$~Gyr for isolated and perturbed disks
(Figure~\ref{fig_Lz-M} and Figure~\ref{fig_Lz-i}) indicate that the
accretion of a satellite may result in the suppression of further
changes of angular momentum of the disk particles . This can be
attributed to the suppression of non-axisymmetric patterns induced by
the central bar after merging (see discussion below).

Thus, the role of satellite accretion in inducing radial migration of
the disk particles strongly depends on the orbit orientation and mass
of the satellite. In general, the combined effects of spiral structure
and bar patterns with satellite-induced migration mechanisms can
produce very sophisticated migration patterns.

\subsection{Fourier analysis}

We calculate the discrete Fourier transform of the bulge+disk surface
density.  We divide the galactic disk in concentric rings and define
the Fourier amplitude coefficient from $m=0$ to $m=4$ in each ring.

\begin{figure*}[htb]
  \resizebox{0.95\hsize}{!}{\includegraphics{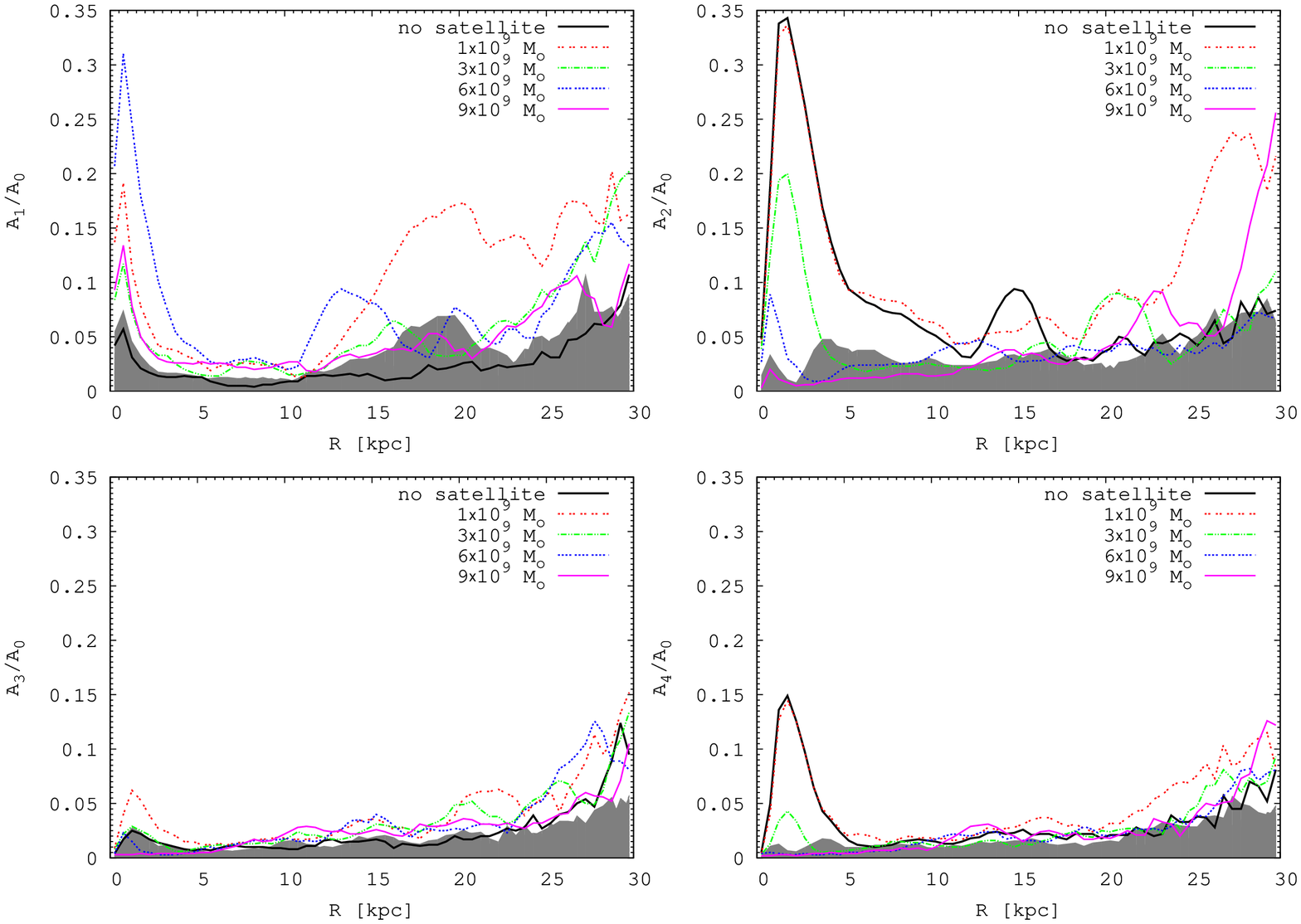}}
  \caption{Fourier amplitudes $A_1$/$A_0$, $A_2$/$A_0$, $A_3$/$A_0$, $A_4$/$A_0$ 
	as a function of radius estimated from the bulge+disk surface density at 4~Gyr 
	for the model A00 (without satellite) and the models A02, B02, C02, D02 (with 
	different satellite masses and fixed satellite initial orbital inclination $i = 30^\circ$).
	The gray regions are estimates of the significance limits
(see description in the text).} 
  \label{fig_fft}
\end{figure*}

Figure~\ref{fig_fft} shows the radial dependence of the mean
Fourier amplitudes $A_1$/$A_0$, $A_2$/$A_0$, $A_3$/$A_0$, and
$A_4$/$A_0$ from the bulge+disk surface density at a time bin
near 4~Gyr (averaged from 19 snapshots at time intervals from 3.8 to
4.2~Gyr) for the model A00 (without satellite) and the models A02,
B02, C02, and D02 (with satellites of different masses). The gray
regions are estimates of the significance limits of the Fourier
amplitudes, i.e., the standard deviation of the values of the Fourier
amplitudes obtained from 19 snapshots at the mentioned time intervals
from 3.8 to 4.2~Gyr. Since the standard deviations of the values of
the Fourier amplitudes for the models A00, A02, B02, C02, and D02 have
comparable radial profiles, we plot only the maximal value of the
standard deviation of the presented models. In all cases halo
and satellite particles were not taken into account. All Fourier
amplitudes $A_{\rm m}$ are normalized by the main axisymmetric
component $A_0$. High amplitudes of $A_2$/$A_0$ in the central disk
region clearly confirm the existence of the central bar. Our
simulations show that the host galaxy bar strength (in the terms of
their Fourier amplitudes $A_2$/$A_0$) becomes weaker for lower-mass
ratios of the merged galaxies.

\section{Conclusions}

We investigate the dynamical interaction and merging of a Milky
Way-like galaxy (host galaxy) with low-mass satellites (mass ratio
1:70 -- 1:8) during the last four Gyr.  We use more than 7.4 million
N-body particles to study the dynamical evolution of such systems (2
million particles for the disk, 0.4 million for the bulge, and 5
million for the halo of the host galaxy and between 33000 to 300000
for a satellite).  We examine the possible changes of the radial
abundance gradient in the disk of the host galaxy caused by particle
migration induced by the merger or interaction. The initial abundance
distribution in the disk of the host galaxy is assumed to be an
exponential (linear in $\log$(O/H) scale). We computed a large set of
mergers with different initial configurations, i.e., with  different
satellite masses, positions, and orbital velocities.  We consider the
evolution of a system over the period of 4 Gyr without star formation. 

We find that there are no significant metallicity changes at any
radius due to  particle migration in the case of the accretion of a
low-mass satellite of $10^9$~M$_{\odot}$ (mass ratio 1:70) except for
the special case of prograde satellite motion in the disk plane of the
host galaxy.  In general, the change in the abundance distribution in
the disk of the host galaxy depends not only on the satellite mass but
also on the initial inclination of the satellite orbit.  The largest
changes take place if the satellite orbit lies in the disk plane of
the host galaxy with prograde motion ($i = 0^\circ$), i.e., the
satellite moves through the disk in the direction of the host galaxy
rotation.    The smallest (if any) changes take place if the satellite
has an initial retrograde orbital motion with an orbital inclination
$i = 90^\circ$ to 150$^\circ$.  Abundances at galactocentric
distances larger than $\sim 10$~kpc can show a significant increase in
the case of accretion of a satellite with a mass $\gtrsim 3 \times
10^9$~M$_{\odot}$ (mass ratio 1:23). 

The merger can result in a change of the slope of the abundance
gradient.  The radial abundance gradient within the central region of
the disk of the host is rather weakly affected by a merger with a
satellite of a mass up to $9 \times 10^9$~M$_{\odot}$. The radial
abundance gradient flattens in the range of galactocentric distances
from 5 to 15~kpc in the case of a merger with a satellite of a mass
$\gtrsim 3 \times 10^9$~M$_{\odot}$.  There is no significant
change in the abundance gradient slope in the outer disk ($\sim$
15~kpc -- 25~kpc) in any merger while the scatter in metallicities at
a given radius increases significantly for most of the satellite
initial masses/positions compared to the isolated galaxy case.
Assuming a mass-to-light ratio in the B band of $M/L_B = 1.4$
\citep{Flynn2006}, the optical ($R_{25}$) radius of our host galaxy is
$\sim$ 15~kpc.
This argues against attributing the break (flattening) of the
abundance gradient near the optical radius observed in the extended
disks of Milky Way-like galaxies exclusively to merger-induced stellar
migration.

\acknowledgments
\section*{Acknowledgments}

We acknowledge the financial support by the Deutsche 
Forschungsgemeinschaft (DFG) through SFB 881 ``The Milky Way System'' 
at the Ruprecht-Karls-Universit\"at Heidelberg, particularly through 
the subprojects A2, A5, and Z2.
IAZ and PB acknowledge also the special support by the NASU under the 
Main Astronomical Observatory GRID/GPU computing cluster {\tt golowood} 
project. 
The main part of the simulations presented here was performed on the 
dedicated GPU clusters {\tt hydra} \& {\tt kepler} at the ARI, funded under 
the grants I/80 041-043 and I/81 396 of the Volkswagen Foundation 
and the grants 823.219-439/30 and /36 of the Ministry of Science, 
Research and the Arts of Baden-W\"urttemberg, Germany.
In addition, part of the code development work was conducted using 
the resources of the GPU cluster {\tt laohu} at the Center of 
Information and Computing at the National Astronomical Observatories, 
Chinese Academy of Sciences, funded by the Ministry of Finance 
of People's Republic of China under the grant ZDYZ2008-2.
PB acknowledges support by the ``Qianren'' (Thousand Talent) 
special foreign experts program of China (Project P.I.\ R.\ Spurzem).
This work was partly funded by the subsidy allocated to Kazan Federal 
University for the state assignment in the sphere of scientific 
activities (L.S.P.).
We are grateful to the referee for his or her constructive comments which 
significantly improved the quality of the paper. 



\bibliography{chem-grad}


\newpage

\newpage

\newpage

\newpage

\newpage

\newpage

\newpage

\newpage

\end{document}